\documentclass[%
 aip,
 amsmath,amssymb,
preprint,%
]{revtex4-1}

\usepackage{graphicx}
\usepackage{dcolumn}
\usepackage{bm}
\graphicspath{./Figures/}
\usepackage[utf8]{inputenc}
\usepackage[T1]{fontenc}
\usepackage{mathptmx}
\usepackage{etoolbox}
\usepackage[dvipsnames]{xcolor}
\usepackage{listings}
\newcommand{\reva}[1]{{{#1}}}
\newcommand{\revb}[1]{{{#1}}}

\usepackage{hyperref}
\usepackage{comment}
\hypersetup{bookmarks=true, colorlinks=true, citecolor=red, linkcolor=blue, urlcolor=blue, linktoc=all, pagebackref, hypertexnames=false}
\makeatletter
\DeclareRobustCommand\citenum
   {\begingroup
    \NAT@swatrue\let\NAT@ctype\z@\NAT@parfalse\let\textsuperscript\relax
     \NAT@citexnum[][]}
\makeatother

\makeatletter
\def\@email#1#2{%
 \endgroup
 \patchcmd{\titleblock@produce}
  {\frontmatter@RRAPformat}
  {\frontmatter@RRAPformat{\produce@RRAP{*#1\href{mailto:#2}{#2}}}\frontmatter@RRAPformat}
  {}{}
}%
\makeatother
\begin{document}

\preprint{LA-UR-24-26289}

\title{Latent Space Dynamics Learning for Stiff Collisional-radiative Models}

\author{Xuping Xie}
\affiliation{Theoretical Division, Los Alamos National Laboratory, Los Alamos, NM 87545, USA}
\affiliation{Department of Mathematics and Statistics, Old Dominion University, Norfolk, VA 23529, USA}
\email[Corresponding author: ]{xxie@odu.edu}

\author{Qi Tang}
\affiliation{Theoretical Division, Los Alamos National Laboratory, Los Alamos, NM 87545, USA}
\affiliation{School of Computational Science and Engineering, Georgia Institute of Technology, Atlanta, GA 30332, USA}

\author{Xianzhu Tang}
\affiliation{Theoretical Division, Los Alamos National Laboratory, Los Alamos, NM 87545, USA}


\begin{abstract}
In this work, we propose a data-driven method to discover the latent space and learn the corresponding latent dynamics for a collisional-radiative (CR) model in radiative plasma simulations. The CR model, consisting of high-dimensional stiff ordinary differential equations (ODEs), must be solved at each grid point in the configuration space, leading to significant computational costs in plasma simulations. Our method employs a physics-assisted autoencoder to extract a low-dimensional latent representation of the original CR system. A flow map neural network is then used to learn the latent dynamics. Once trained, the reduced surrogate model predicts the entire latent dynamics given only the initial condition by iteratively applying the flow map. The radiative power loss is then reconstructed using a decoder. Numerical experiments demonstrate that the proposed architecture can accurately predict both the full-order CR dynamics and the radiative power loss rate.
\end{abstract}

\maketitle


\section{\label{sec:intro} Introduction}

Collisional-radiative (CR) models describe the atomic processes in a
plasma by tracking the population density in the ground and excited
states for each charge state of the atom or ion. {These models predict
important plasma properties such as charge state distributions and
radiative emissivity and opacity. CR models play a crucial role in understanding and predicting the behavior of plasmas in various fields, including gas discharges, aerospace applications, and astrophysics, see Ref.~[\citenum{capitelli2012fundamental,arnaud1992iron,lieberman1994principles,celiberto2017elementary}]. Accurate CR modeling is essential in
radiative plasma modeling for magnetic fusion, especially when
significant amount of impurities are introduced into the plasmas. }

Fusion power reactors, such as tokamaks and stellerators, confine
plasma using magnetic fields to sustain nuclear fusion reactions at
ion temperature of 10-15~keV
and plasma density in the order of $10^{20}\,$m$^{-3}.$ Such high
temperature plasma would eventually transition to a boundary plasma
that is of a few eV next to the divertor plates. Line emission of a
hydrogenic plasma of deuterium and tritium, mixed with helium ash, only becomes a significant factor
at the low-temperature boundary plasma. In actual power reactor
scenarios, seed impurities such as neon and argon are deliberately
introduced into the plasma to radiative strongly at much higher
electron temperature, up to hundreds of eV and even multiple keV. Wall
impurities will also be inevitably brought into the plasma due to
plasma-wall interaction. For the current ITER and various DEMO designs
with solid first wall and divertor, one would have tungsten impurity,
while for a liquid first wall solution, lithium impurity would be
present in the plasma.

The modeling of a radiative plasma needs information on the rate of
change for the ion charge states and the radiative power loss rate,
which are readily available from the solution of a time-dependent
collisional-radiative (CR) model. If the plasma evolves on time scale
much longer than that of the collisional-radiative processes,
steady-state CR results can be coupled to plasma simulation in the
form of tabulated data table. There are important plasma dynamical
phenomena, for example, in tokamak disruptions, that can have very fast
plasma dynamics such as plasma thermal quench. Here a dynamical CR model
is needed for physics fidelity.

Formally the collisional-radiative processes are captured in the
plasma model via the Boltzmann operator that accounts for these
inelastic collisions, in addition to the usual Coulomb collision
operator for the elastic collisions between charged particles. This
full system of partial differential equations (PDE) can be numerically
treated with an operator splitting method that seperately treats the
elastic Coulomb collision operator and the Boltzmann operator for the
collisional-radiative processes. With all the plasma transport terms
lumped with the Coulomb collision operator, the Boltzmann operator
becomes stand-alone and gives rise to the collisional-radiative model
as is commonly known. The coupling between the CR model and the plasma
transport model can be formally implemented through the operator
splitting scheme.  The simplest coupling problem is for a homogeneous
plasma, in which the plasma dependence of the CR model is through
$n_e$ and $T_e.$ Self-consistent evolution of the plasma can be
provided by an energy equation for $T_e$ and the quasineutrality constraint for
$n_e.$

The CR model for fusion plasma must account for a variety of
collisional-radiative processes, for an example, see
Ref.~[\citenum{garland2020impact}].  It is mathematically described by
a high-dimensional nonlinear dynamical system that is very stiff
because of the fast transition rate of ionization, recombination,
excitation, and de-excitation. In practice, high-fidelity simulations
of time-dependent CR model are extremely computationally expensive for
direct coupling with plasma simulations. As a result, there is an
increasing interest and demand for efficient and accurate surrogate
models {from deep learning} for the CR system in plasma disruption
simulations~[\citenum{garland2020progress, garland2022efficient}].

{Although the ultimate goal of this line of research is to couple the CR
model to dynamical plasma modeling, which requires self-consistent
evolution of $n_e$ and $T_e,$ the perceived difficulties in the realm
of deep learning, are already extreme even for a standalone CR model,
compared with what have been done in the literature for time-dependent
ordinary differential equations (ODE). The most prominent challenge is
the many orders of magnitude variation in state variables of the CR
model and the critical physics quantity of radiative power loss rate,
which pose an extreme challenge in accuracy for deep learning.  Another prominent
challenge is related to the eventual goal of coupling the CR surrogate
to the plasma transport model, which requires a deep neural network (DNN) surrogate that can
advance the CR dynamics over a time step $\Delta t.$ The stiffness of
the CR model introduces an exponentially varying $\Delta t$ for its
temporal integration. The DNN surrogate for the one-time-step forward
map of CR system of equations, thus must be trained and remain
accurate over the extreme variation in $\Delta t.$ Both of these
challenges in the realm of deep learning are retained in the
standalone CR model, so in the current work, we take on the limited
scope of developing DNN surrogates for standalone CR models with the
simplification of constant $T_e,$ although the surrogate model will be
trained over a range of $T_e$s.}

Specifically, we will
first use a physics-assisted autoencoder on the CR data to find a
low-dimensional latent representation of the original CR system. Then,
we use a flow map neural network to learn the latent dynamics. Once
our reduced surrogate model is trained, we can predict the whole
latent dynamics given only the initial condition, through iteratively
applying the flow map neural network, and then reconstruct its
radiative power loss via a decoder. By leveraging deep learning
techniques, our proposed surrogate model can provide a computationally
efficient and accurate representation of the CR dynamics, facilitating
better prediction and mitigation of plasma disruptions in fusion
reactors.

\section{\label{sec:CRmodeling}Collisional-radiative Modeling}

\begin{figure}[htp]
    \centering
    \includegraphics[width=0.9\linewidth]{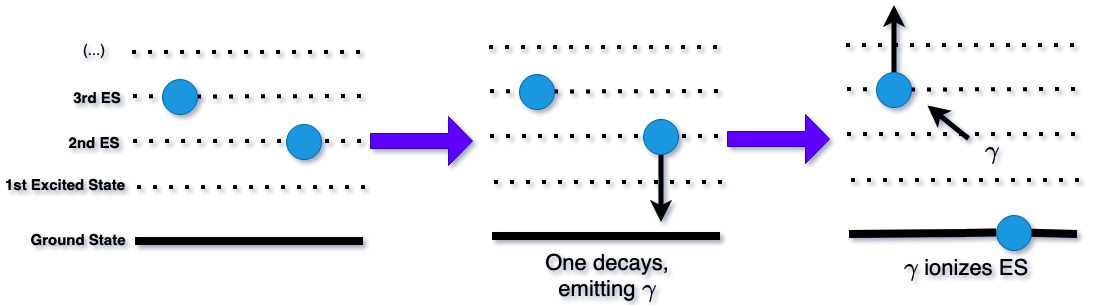}
    \caption{Ion particles ionization and de-excitation in a CR model.}
    \label{cr_demo}
\end{figure}

CR modeling deals with the complex interactions between
electrons/photons and ions/neutrals in a plasma. These interactions
could result in excitation, de-excitation, ionization, and
recombination.  Fig.~\ref{cr_demo} illustrates the processes of
spontaneous decay and photo-ionization of an atom that is initially in
excited states. The goal of CR modeling is to calculate the population
densities of various charge states that resolve both the ground and excited energy levels of neutral and partially ionized atoms in the
plasma, accounting for both collisional and radiative processes. To
manage the wide range of possible ion states in a
plasma, we order the ion population density vector in a plasma as,
\begin{equation*}
    \mathbf{N}=\{N^j_{\alpha,Z}\},
\end{equation*}
where $N^j_{\alpha,Z}$ denotes the population density of ion level $j$
for atomic species $\alpha$ and charge state $Z$, which ranges from 0
to $A_\alpha$, the atomic number of species $\alpha$. The index $j$
labels each of the ground and excited states. Typical impurity species involved in a fusion plasma
include lithium
($A_\alpha=3$), nitrogen ($A_\alpha=7$), neon ($A_\alpha=10$), and
argon ($A_\alpha=18$). The high fidelity solution of a CR model provides critical input into the plasma simulation code.
These include the species ion charge state distribution $n_{\alpha,
  Z}$ and the radiative cooling rate $\mathcal{R}_L$. The species charge state density
$n_{\alpha, Z}$ is simply the sum over all ground and excited states
of a given charge state,
\begin{equation*}
    n_{\alpha,Z} = \sum_{j}N^j_{\alpha, Z}.
\end{equation*}
The radiative power loss (RPL) rate is a crucial quantity
chracterizing the radiative cooling rate of electrons in a plasma.
It can be calculated by summing up
the contributions from all radiative transitions, including line
emissions (bound-bound), recombination radiation (free-bound), and bremsstrahlung (free-free).
As an example, the line emission rate takes the form:
\begin{equation*}
\mathcal{R}_L = \sum_{\alpha,Z,j,k}N_{\alpha,Z}^jA^j_{\alpha,Z\rightarrow k}E^j_{\alpha, Z\rightarrow k},
\end{equation*}
where $A^j_{\alpha,Z\rightarrow k}$ is the coefficient for spontaneous
emission from energy level $j$ to level $k$, and $E^j_{\alpha,
  Z\rightarrow k}$ is the the energy of the photon emitted in the
transition from level $j$ to level $k$.

The CR model is mathematically represented as a parameterized ODE:
\begin{equation}
\frac{d\mathbf{N}}{dt} = {R}(\mathbf{N}; n_A,T_e) \, \mathbf{N},
\label{eqn:rate-matrix}
\end{equation}
where $\mathbf{N}$ is the population density vector of ions in various
charge states including both ground and excites states, $T_e$ is the
temperature, $n_{A, \alpha}=\sum_{Z}n_{\alpha, Z}$ is the total atomic
density for species $\alpha$.  The dimension of $\mathbf{N}$ can vary
enormously depending on whether one would want to resolve the fine and
super-fine structures in the excited states. As an example, LANL's
ATOMIC model [\citenum{hakel2006new}] can have more than $10^6$ states
for argon. The rate matrix ${R}(\mathbf{N}; n_A, T_e)$ is a square
matrix describing a range of atomic processes, which can be broadly grouped
in up-transitions and down-transitions.  {Collisional charge exchange can provide both down-
and up-transitions depending on the specific ion/atom involved. The
overall rate matrix ${R}$ has explicit dependence on the electron
distribution. The common assumption is to approximate the electron
distribution as a Maxwellian with temperature $T_e$ and density $n_e$. There are some works not using this approximation where the electron energy distribution is calculated together with the species density and level distributions, see Ref.~[\citenum{colonna2001coupled,capitelli2012fundamental,capitelli2016particle}].} 
In a quasineutral plasma, the electron density is approximately equal
to ion charge density $\sum_{Z,\alpha} Z n_{\alpha,Z}.$
The species atomic number density $n_\alpha$ or more explicitly labeled as $n_{A,\alpha},$ is the sum of all charge states
$\sum_Z n_{\alpha,Z}.$ for species $\alpha$.

\begin{figure}[!htb]
\centering
        \includegraphics[width=0.8\linewidth]{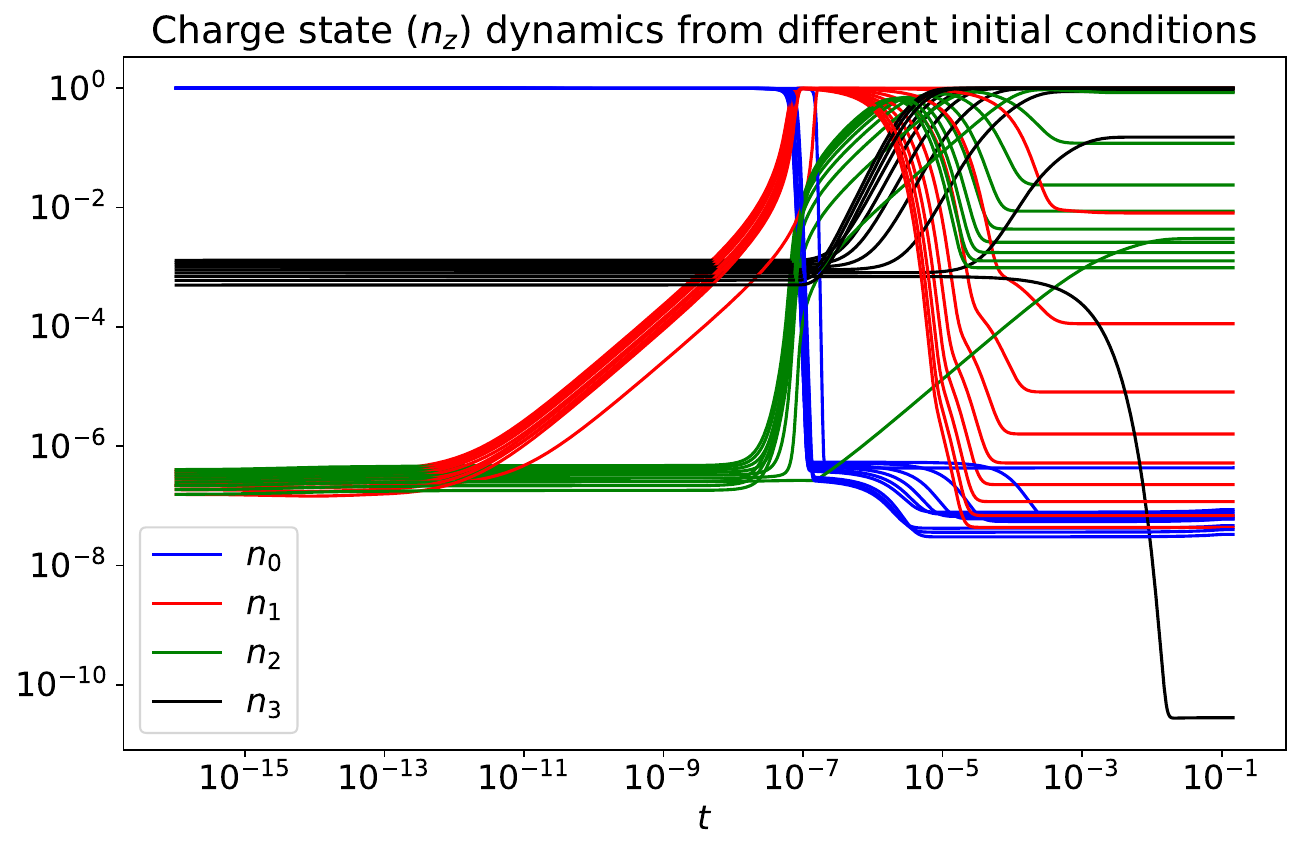}
    \caption{\reva{Trajectory of the normalized charge state from the high-fidelity numerical simulation of the CR model for different initial conditions. The solution is for a single species, lithium, where the atomic number $A_\alpha=3$. Four colored lines represent four charge states (they are partial sums of the solution state vector; details will be defined later), $n_0, n_1, n_2, n_3$. The physical unit of the ion and charge states is $m^{-3}$. The plot shows the non-dimensional form of the charge states after normalization. Refer to Section~\ref{subsec:dataprocessing} for details on the data processing. The horizontal axis represents the numerical simulation time in its non-dimensional form.}}
       \label{fig:data_solution}
\end{figure}

The CR model is a set of parameterized stiff ODE, as many of the transition
rates are very fast compared with the time scale for reaching ionization balance. Numerical solutions of
the full system~\eqref{eqn:rate-matrix} require implicit time
stepping which is usually performed along with a standard linear
algebra package. {Resolving the CR model can be a
computational burden for many degrees of
freedom in $\mathbf{N},$ in which case it would be simply
impossible to couple the CR physics module in its native
form with 3D plasma simulations, since for each spatial grid point in
the plasma simulation, one would need to evolve a coupled ODE with
many degrees of freedom, which would overwhelm the plasma
simulation cost by many orders of magnitude. In our numerical experiment, as a proof-of-concept demonstration, the system contains 94 states using LANL's FLYCHK code. A similar numerical implementation can be found in Ref.~[\citenum{ninni2022influence}]}. The goal of this work is
to introduce a reduced-order model that can efficiently and accurately
predict the fast dynamical transition of the charge state,
$n_{\alpha,Z}$, and the radiative cooling rate, $\mathcal{R}_L$.

\subsection{\label{ROM}Data-driven reduced-order Modeling}

Reduced Order Modeling (ROM) seeks low-dimensional approximations of
high-dimensional systems, in order to significantly reduce computational
costs. In the framework of data-driven ROM, machine learning (ML)
techniques and data are used to distill the essential features of
complex systems into more manageable representations. This is
especially relevant in fields such as fluid dynamics, material
science, and climate modeling, where solving full-scale problems
involves significant computational challenges due to high
dimensionality. Proper Orthogonal Decomposition (POD) and Dynamic Mode
Decomposition (DMD) are two of the most popular approaches to extract
dominant features and dynamic structures from data. These methods have
been successfully applied in fluid mechanics, control, and
biomechanical problems~[\citenum{xie2018data, xie2017approximate,
    amsallem2012stabilization, snyder2023numerical,
    peherstorfer2015dynamic}]. With the development of deep learning,
autoencoder neural networks have gained significant attention in ROM
due to their ability to efficiently compress high-dimensional data
into a lower-dimensional latent space while preserving essential
features.  Recent works have demonstrated the use of autoencoders for
learning low-dimensional representations of fluid
dynamics~[\citenum{mardt2018vampnets,
    hasegawa2020machine}]. Ref.~[\citenum{champion2019data}] combined
autoencoders with DMD to learn the governing equations of dynamical
systems directly from data, allowing for efficient predictions of
future states.

Data-driven discovery of dynamics has emerged as a powerful tool for
modeling and predicting the behavior of complex physical systems. This
approach leverages ML techniques to extract patterns and underlying
dynamics from data, facilitating the development of reduced-order
models that are both efficient and
accurate. Ref.~[\citenum{brunton2016discovering}] introduced the
Sparse Identification of Nonlinear Dynamical Systems (SINDy) method,
which aims to discover governing equations from data. By representing
the dynamics in a library of candidate functions, SINDy selects the
most relevant terms to construct a parsimonious model. This
data-driven methodology has proven effective in capturing the
underlying physics of complex systems with minimal assumptions,
offering a powerful tool for system identification and model
reduction~[\citenum{fukami2021sparse,
    kaheman2020sindy}]. 
Ref.~[\citenum{chen2018neural}] introduced
Neural Ordinary Differential Equations (NODE), which parametrize the
time derivative of the hidden state with a neural network, allowing
the model to learn complex dynamics directly from data. This method
can also be applied for identifying latent
dynamics~[\citenum{rubanova2019latent,
    linot2023stabilized}]. Ref.~[\citenum{lusch2018deep}] utilized
autoencoders to identify a latent space where the nonlinear dynamics
are approximated by linear models based on linear Koopman operator
theory, enabling efficient and accurate predictions of system
behavior. Ref.~[\citenum{koronaki2024nonlinear}] introduced physics-
and data-assisted ROM based on approximate inertial manifolds theory
using deep neural networks. Their approach is successfully
demonstrated through dissipative
PDEs. Ref.~[\citenum{koronaki2024nonlinear}] discussed the
``physics-assisted'' latent space learning where the latent space is a
``grey-box'' approach including the known physics latent space and
data-driven latent space. A few recent works on learning flow maps
using structure-preserving neural networks can be found
in~[\citenum{burby2020fast, duruisseaux2023approximation}], which has been recently applied to learning beam dynamics in accelerators [\citenum{huang2024symplectic}]. These
studies underscore the potential of integrating ROM approach with
latent dynamics learning to address the computational challenges in
simulating high-dimensional dynamical systems.

\subsection{Specific aim and approach of this work}

The main challenge in finding the reduced CR model for coupling to
plasma simulations lies in two parts. The first is to efficiently
identify the low-dimensional representation from the data. The second
is to learn latent dynamics that can accurately predict the
trajectories of the reduced system using only the initial
conditions. This allows for an approximation of the full-order
dynamics and the radiative cooling rate with high
accuracy. Fig.~\ref{fig:data_solution} plots the trajectories of
different charge states from the high-fidelity numerical
simulation. It is evident that the sharp transitions over very short
time scales present a significant challenge in accurately modeling the
reduced CR dynamics.

It should be emphasized that for the purpose of coupling the surrogate
CR model to plasma simulations, the two essential quantities, which
are of explicit physics meaning, are the species ion charge state
density $n_{\alpha,Z}(t)$ and the radiative power loss (RPL) rate
$\mathcal{R}_L(t).$ \revb{The first will enter the plasma model to update
the ion charge distribution, while the second will enter the plasma
model as an energy sink term for the electron thermal energy.  Since
we have decided to evolve a surrogate model in a much lower
dimensional latent space to save computational cost in coupled
CR-plasma simulations, it becomes a necessity to include the physical
variables $n_{\alpha,Z}$ in the latent space.  This naturally leads to
a grey box latent space in which physical variables in the white space
and the usual blackbox variables that have no physical meaning, would
co-reside.  So the actual latent space discovery, which refers to the
unknown black space variables, is constrained by the required presence
of white space variables for the decoder network and its training.} In
other words, the deployment of white space variable in the grey box
latent space is required to facilitate the coupling of the latent
space surrogate model to plasma simulations. It is unclear, or at
least we do not know, if the presence of these white space variables
is actually beneficial to extract a more optimal black subspace, in
terms of both the training cost for the encoder-decoder and the
minimal size of the latent space. For potential benefit of
``physics-assisted'' as opposed to ``physics-constrained'' latent space
discovery reported here, one can consult
Ref.~[\citenum{koronaki2024nonlinear}].

Consistent with the need of direct coupling to the plasma simulation
model, the radiative power loss rate $\mathcal{R}_L$ will be
reconstructed with the decoder, in addition to the full vector of
$\mathbf{N}.$ The CR time dynamics, from the data, is given by
$\mathbf{N}(t_n)$ at discrete time $t_n,$ with a variable time step
$\Delta t_n = t_{n+1} - t_n.$ The corresponding latent space dynamics
is modeled by a flow-map neutral netwrrk, which is a common approach
for learning discrete dynamical system  [\citenum{rnn, lstm}].
It is interesting to note that
the extreme variation in $\Delta t_n$ makes our dynamical surrogate
reconstruction a standout among those flow-map NN work reported in the
literature.  For the purpose of demonstration, our approach will be
 tested by the CR data from a single species, lithium,
under different parameters and initial conditions in Section
\ref{sec:numeric}.

\section{Data-Driven Model Reduction for CR modeling}
\label{sec:ddrom}
\begin{figure}[htp]
    \centering
    \includegraphics[width=0.9\linewidth]{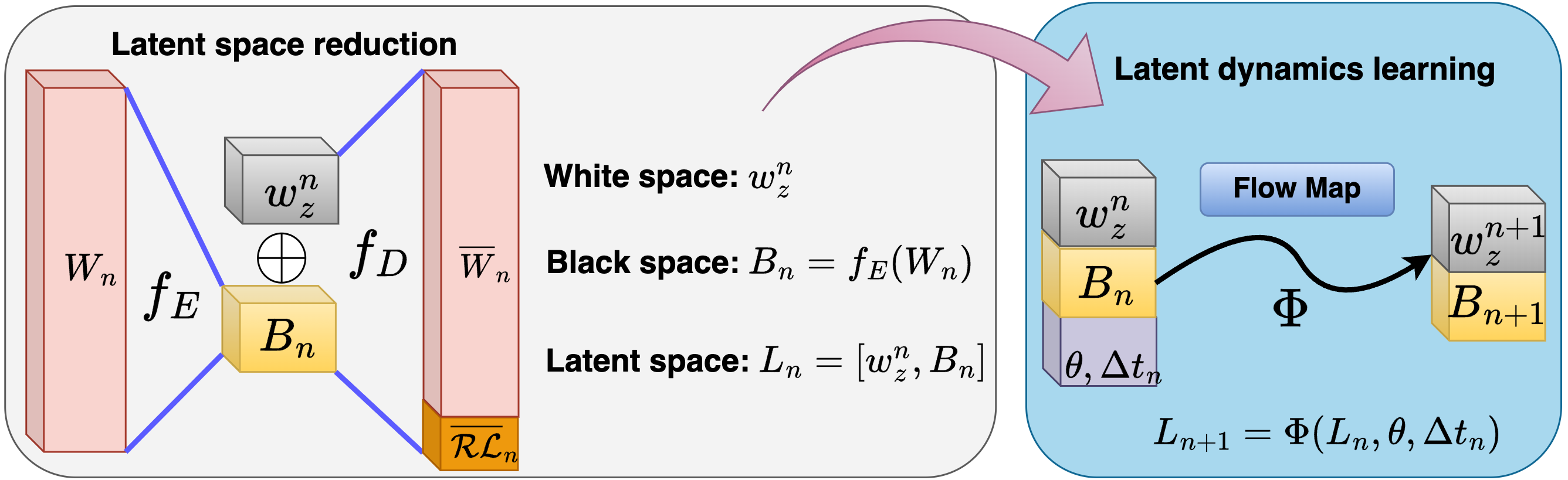}
    \caption{Autoencoder architecture for physics-assisted latent space identification (left). Flow map for latent dynamics learning (right). $\theta$ denotes the parameter space in our CR system which are total density, $n_A$, and temperature, $T_e$.
    }
    \label{fig:rpm_demo}
\end{figure}

This section describes the deep learning approach for 
our latent space model reduction of the CR data. The
methodology involves two crucial steps: latent space discovery (reduction) and
latent dynamics learning, both of which are facilitated by designing
an appropriate neural network architecture. The network comprises an
encoder to reduce the input high dimensional data into a lower-dimension latent space,
followed by a neural network to learn the flow map of the latent
space dynamics. Subsequently, a decoder is employed to reconstruct the full ion
density space and radiative power loss rate $\mathcal{R}_L.$ Fig.~\ref{fig:rpm_demo} illustrates the architecture of
our data-driven ML-based surrogate for the CR system.

\subsection{Grey Latent Space Discovery via Autoencoder}

{We first introduce the concepts of white space, black space, and grey box to describe different components of the latent space—the reduced representation of the original high-dimensional system. \textbf{White space} refers to the part of the latent space composed of physical quantities that are directly interpretable and have well-established meanings within the CR model. We include the charge states ($n_{\alpha,Z}$) of the plasma explicitly in the latent space as white space variables. These are known physical parameters that play a crucial role in plasma behavior. By incorporating these known quantities, we ensure that essential physical properties are retained in the reduced model, making this part of the model transparent and easily understandable. \textbf{Black space} represents the portion of the latent space learned by the autoencoder without direct physical interpretation. It consists of abstract features extracted from the data. The autoencoder compresses the remaining high-dimensional information into a lower-dimensional representation in the black space. These dimensions capture complex patterns and relationships not easily described by known physical variables. This component functions as a black box since the internal representations are not directly interpretable. It allows the model to capture intricate dynamics and nonlinearities inherent in the CR system. The \textbf{Grey box} refers to the combined latent space that integrates both white space (interpretable physical quantities) and black space (learned abstract features). Our overall latent space is a grey box because it merges the physically interpretable charge states with the abstract features learned by the autoencoder.}

Autoencoders consist of two main components: an encoder function
$f_E$, which maps the input data to a lower-dimensional latent space,
and a decoder function $f_D$, which reconstructs the original data
from its latent representation. During the training, the autoencoder
minimizes the reconstruction error between the input and the
reconstructed output, thereby capturing the most significant features
of the data. In the first part of our data-driven framework, an
autoencoder is utilized to reduce the full ion population space into a
black space, $B_n$, at a given time step $t_n$. The black
space in combination with the known physics information of the charge
state (i.e., a white space, $w_z^n$) forms our latent space,
$L_n$. The full ion population can then be reconstructed through a
decoder $f_D$ from this latent space. Note that when setting the
dimension of the black space to zero, we recover the charge states as
our latent space, which would be the limit of a time-dependent coronal
equilibrium model. This is usually too crude a reduction to recover
accurately the radiative power loss rate $\mathcal{R}_L,$ which we shall retrieve
from the decoder, through a finite-sized black subspace in the
autoencoder latent space.  In our case, the encoder part transforms
the input, \revb{i.e., a normalized ion population \( W_n \in \mathbb{R}^{N}
\), to a black space \( B_n \in \mathbb{R}^{r} \) with \( r \ll N \):}
\begin{equation*}
B_n = f_E(W_n).
\end{equation*}
The decoder part tries to reconstruct the input \( W_n \) and predict
the radiative loss from the latent space \( L_n=[w_z^n, B_n] \):
\begin{align*}
[\overline{W_n}, \overline{\mathcal{R}_{L, n}}] = f_D(L_n).
\end{align*}

The objective is to minimize the reconstruction loss \(
\mathcal{L}_{recons} \), which measures the difference between the
input \( W_n \) and the reconstructed output \( \overline{W_n} \) and
\(\overline{\mathcal{R}_{L, n}}\):
\begin{equation*}
\mathcal{L}_{\rm recons} = \frac{1}{S}\sum_{n=1}^S \Big[ \|W_n -
  \overline{W_n}\|^2 + \|\mathcal{R}_{L, n}-\overline{\mathcal{R}_{L,
      n}}\|^2 \Big],
\end{equation*}
where $S$ is the total number of time steps in our training dataset. 


\subsection{Latent Dynamics Learning via Flow Maps}

A dynamical CR model is formally a
continuous-time dynamical systems described by ordinary differential
equations (ODEs). Its latent system's behavior can be expressed as:
\begin{equation*}
\frac{d\mathbf{L}}{dt} = \mathbf{F}(\mathbf{L; \theta}),
\end{equation*}
Here, \(\mathbf{L} \in \mathbb{R}^r\) is the reduced latent state
vector, and \(\mathbf{F}(\mathbf{L}; \theta)\) is the vector field
that describes the dynamics of the system parameterized by $\theta$.
One can construct such a system and pursue a neural ODE integration scheme
for coupling the surrogate CR model to plasma simulation.

In this work, we opt to construct a surrogate CR model in the form of a discrete-time
flow map from $t$ to $t+\Delta t,$ which describes the 
time integration of the ODE system over discrete time steps.
The flow map \(\Phi(\Delta t, \mathbf{L}_0)\), is a function that maps
the initial state \(\mathbf{L}_0\) at \(t=t_0\) to its future state
\(\mathbf{L}(t)\) at a time interval \(\Delta t\), such that:
\begin{equation*}
\mathbf{L}(t_0 + \Delta t) = \Phi(\Delta t, \mathbf{L}(t_0))
\end{equation*}
This discrete-time flow map arises naturally in coupling the CR model
with plasma simulations using a finite step size $\Delta t$ in time
integration.  In general, for discrete-time dynamical systems with a
uniform time step, the system can often be described by:
\begin{equation*}
\mathbf{L}_{n+1} = \mathbf{G}(\mathbf{L}_n).
\end{equation*}
Since time integration of our CR model has exponentially varying time steps and is also
parameterized by the total density $n_A$ and temperature $T_e$, our
flow map \(\Phi\) acts at discrete steps, mapping \(\mathbf{L}_n\) to
\(\mathbf{L}_{n+1}\) and can be denoted as:
\begin{equation*}
\mathbf{L}_{n+1} = \Phi(\Delta t_n, \mathbf{L}_n; n_A, T_e),
\end{equation*}
see the Fig.~\ref{fig:rpm_demo} for the model architecture. In this
case $\theta$ includes $n_A$ and $T_e$.  Thus, the flow map can be
approximated by a neural network with the loss function defined as the
following,
\begin{equation*}
\mathcal{L} = \frac{1}{S}\sum_{n=0}^S \Big[\|\mathbf{L}_{n+1} -
  \Phi(\Delta t_n, \mathbf{L}_n; \theta)\|^2+\|n_A-\sum_{i=0}^Z
  n^{n+1}_i\|^2\Big],
\end{equation*}
where $\Phi(\Delta t_n, \mathbf{L}_n; \theta)$ is the neural network
prediction of the trajectory at next time step, and
$\|n_A-\sum_{i=0}^Z n^{n+1}_i\|^2$ is the additional physical
constraint, i.e., the mass conservation of atomic species.

\section{Numerical Experiment}
\label{sec:numeric}

In this section, we present detailed numerical experiments for the
proposed surrogate model in Section~\ref{sec:ddrom}.
As a proof of concept, the case study is performed with parametrized CR model with a single atomic species of lithium, using
a superconfiguration atomic model and the rates from the LANL FLYCHK code [\citenum{chung2005flychk}]. 

\subsection{\label{subsec:dataprocessing} Data processing}

The success of the ML training heavily relies on the proper data processing. 
Here we describe two critical components to rescale the data to ease the ML training.
A proper sampling for stiff dynamics is also an important step to guarantee a good training result.

\begin{figure}[!htb]
\centering
        \includegraphics[width=1\linewidth]{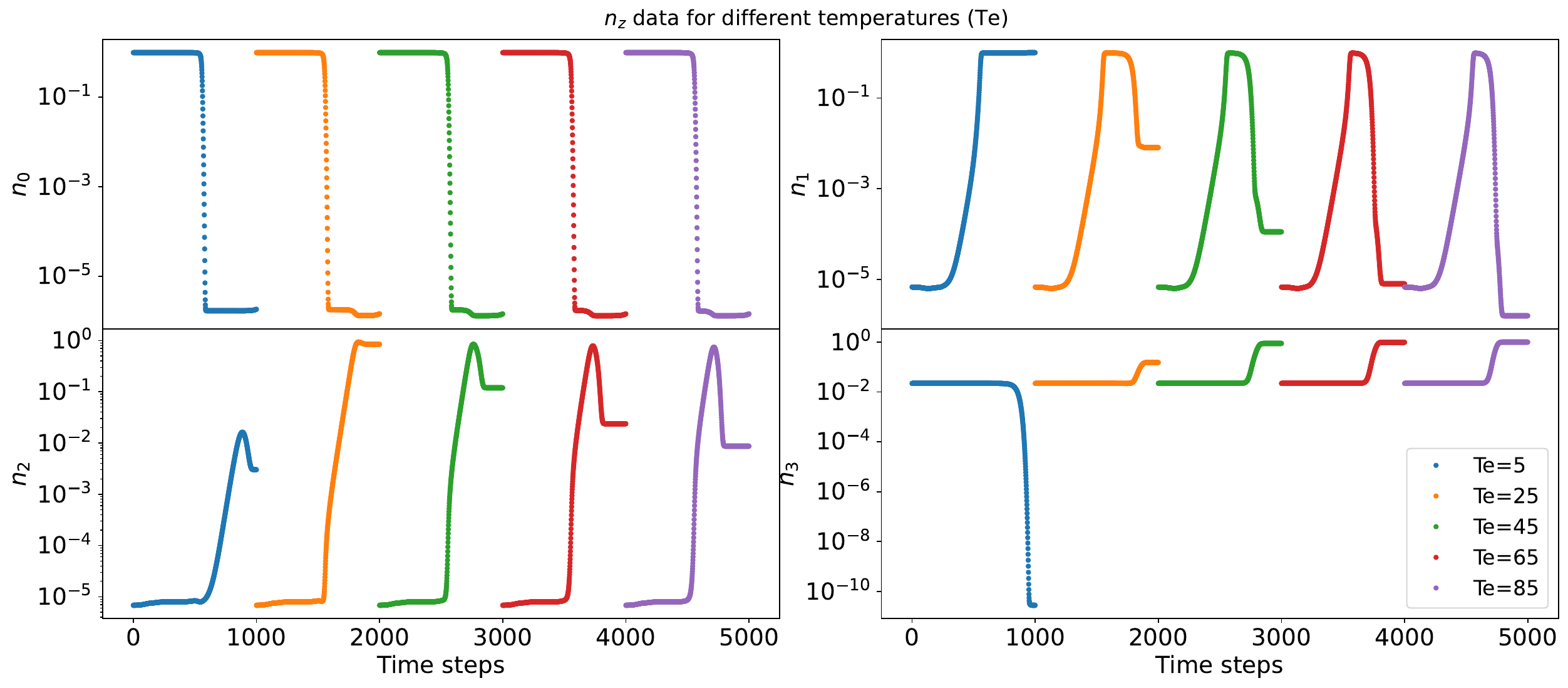}
    \caption{\reva{High-fidelity CR solution. Normalized charge state trajectories from 5 different temperatures $T_e$ are plotted. All trajectories share the same X-axis (time steps), and each trajectory corresponds to a separate simulation for the respective temperature.}}
    \label{fig:4ktraj_Te}
\end{figure}

\paragraph{\textbf{Ion Density Normalization.}}  \revb{The CR solution of Lithium has
$\dim{\mathbf{N}}=94$ states corresponding to different energy levels with the atomic number $A_\alpha=3$. The magnitude of ion population $\mathbf{N}$ from the numerical solution varies largely between 1e13 to 1e-11 ($m^{-3}$). Since the total density $n_A$ is preserved for different parameter setting, to facilitate the neural network training process, we first normalize the population using its total density $n_A$ as the follows,
\begin{equation*}
\tilde{\mathbf{N}} = \frac{\mathbf{N}}{n_A}, \qquad n_A=\sum_{i=0}^{94} \mathbf{N}_i
\end{equation*}
The corresponding normalized charges are, 
\begin{equation*}
{n}_0 := \sum_{i=0}^{31} \tilde{\mathbf{N}}_i, \qquad
n_1 := \sum_{i=32}^{62} \tilde{\mathbf{N}}_i, \qquad
n_2 := \sum_{i=63}^{93} \tilde{\mathbf{N}}_i, \qquad
n_3 := \tilde{\mathbf{N}}_{94}.
\end{equation*}
The physical units of the charge states ($n_z$) and corresponding radiative loss rate ($\mathcal{R}_L$) are $m^{-3}$ and $m^{-3}/s$. After applying total density scaling, the range of ion population density $\tilde{\mathbf{N}}$ lies in $($1e-27$,1)$. Additionally, the quantities of interest, i.e., $n_z$ and $\mathcal{R}_L$ are presented in a non-dimensional form in a ll plots.  Fig.~\ref{fig:data_solution} shows the charge state trajectories from different initial conditions from the high-fidelity CR model. We can clearly see that there is sharp transition at very tiny time scale in the CR dynamics, which makes it one of many challenges in the reduced latent dynamics learning. 
Many of the ion densities are still very small, with magnitudes less than 1e-10, making it challenging for the neural network to learn such values of tiny magnitude.  To accurately resolve the high-fidelity dynamics of the CR model, it is important to resolve small values of the states.} \reva{In Fig. \ref{fig:4ktraj_Te}, we display the trajectories of the charge states for different electron temperatures ($T_e=5,25,45,65,85$). All trajectories are plotted over the same X-axis, which represents time steps. Each trajectory corresponds to a separate simulation run with a different temperature $T_e$, but all simulations run for the same number of time steps. There is no shifting in time between the trajectories; they are just presented together on the same plot for easy comparison. This allows us to observe how the different temperature values influence the evolution of the charge states over time. The shared X-axis facilitates visual comparison, while the vertical axis (logarithmic scale) captures the behavior of the charge states over time for each temperature.} 

\begin{figure}[!htb]
\centering
        \includegraphics[width=0.48\linewidth]{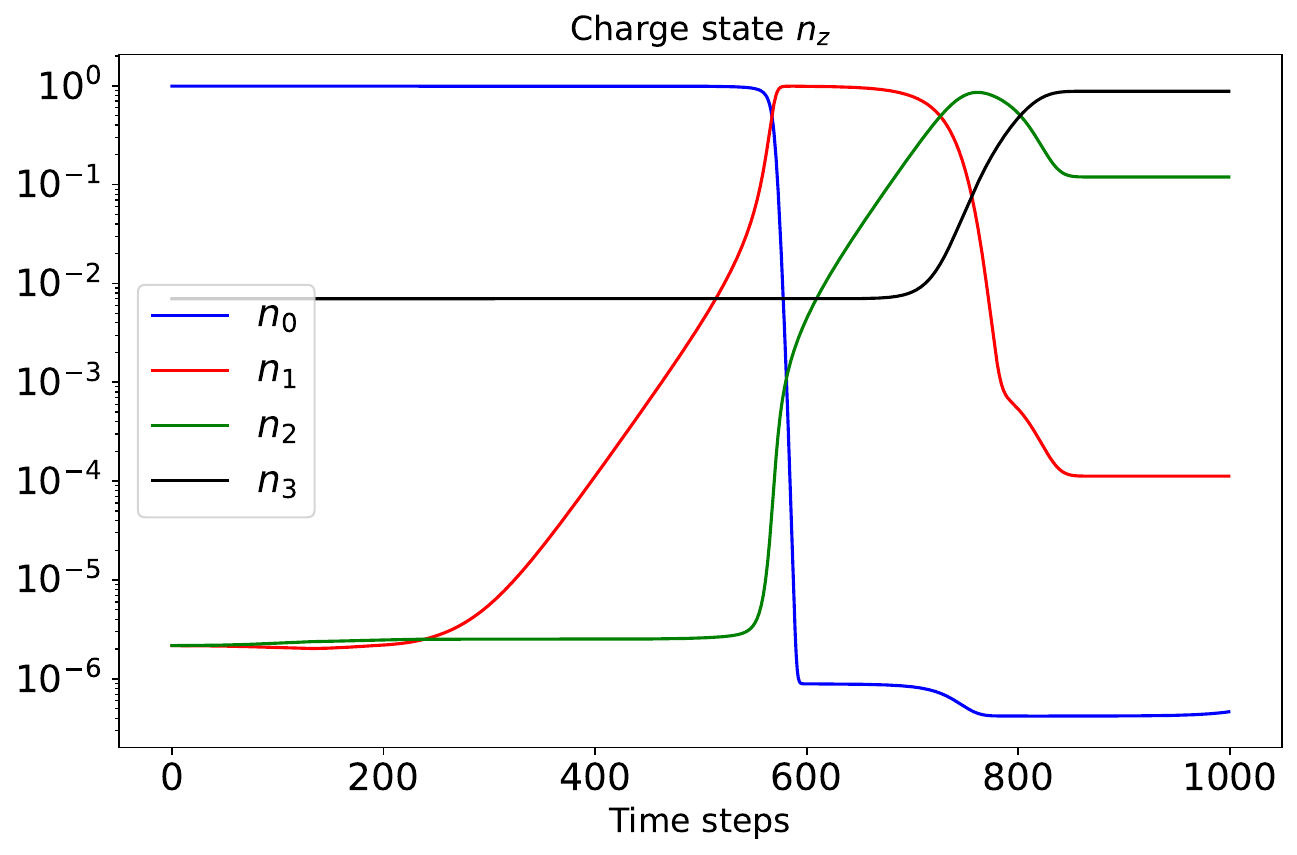}
        \includegraphics[width=0.48\linewidth]{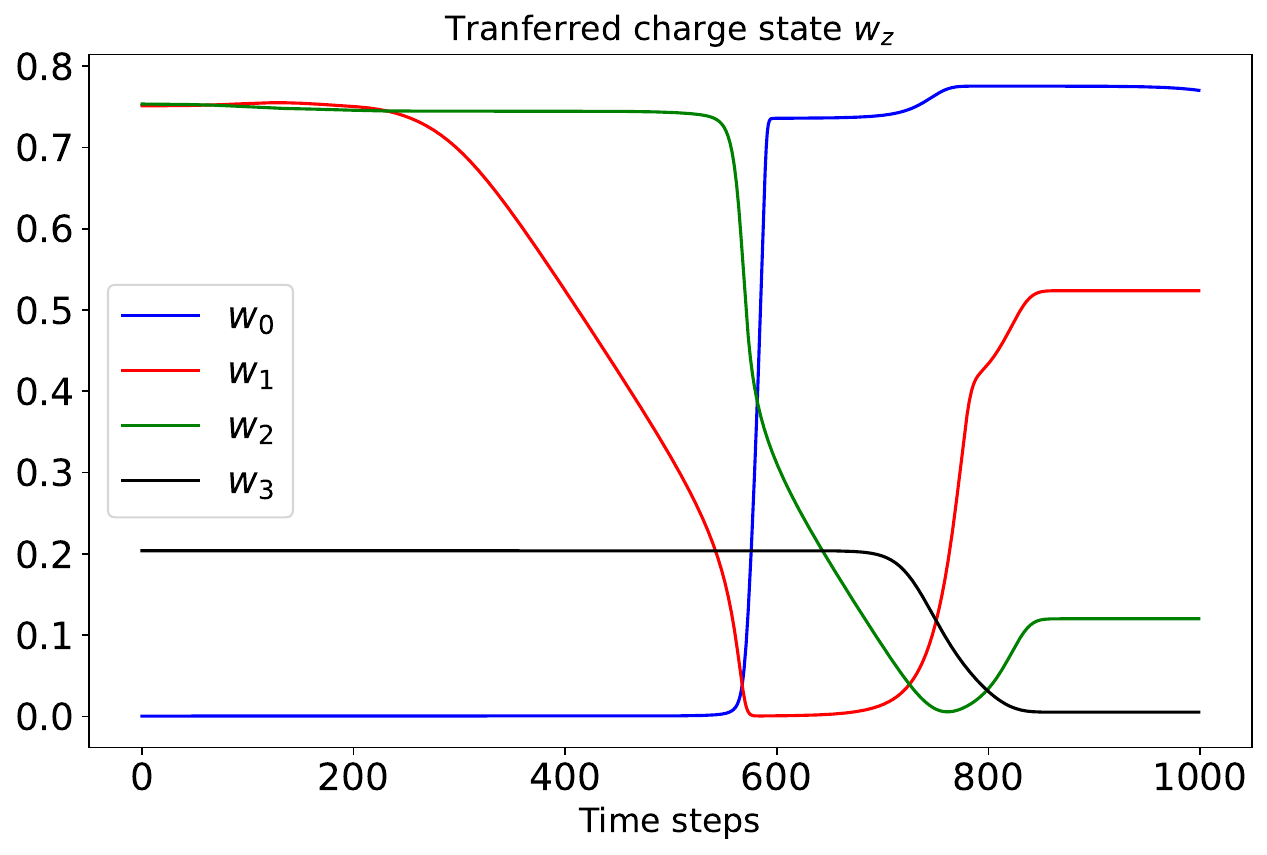}
    \caption{Charge states in the original scale, $n_z$, and in the transferred scale, $w_z$.
    The latter is used in the neural network training. }
    \label{fig:min_max_w}
\end{figure}

To accommodate that, we first apply the following change of variable to transform the data into a more suitable range:
\begin{equation*}
\tilde{W} = {1-\log(\tilde{\mathbf{N}})},
\end{equation*}
then using $(0,1)$ min-max scaler on $\tilde{W}$ to obtain the properly scaled training data $W$:
\begin{equation*}
    W=\frac{\tilde{W}-\tilde{W}_{min}}{\tilde{W}_{max}-\tilde{W}_{min}}.
\end{equation*}
The transformed variable $W$ lies in $(0, 1)$ and is properly scaled, making it easier for the neural network to train effectively.  Fig.~\ref{fig:min_max_w} shows the charge state, $n_z$, both in its original scale and its normalized scale (between 0 and 1), as used in practical neural network training. Note that we use the notation $w_z$ (w0, w1, w2, w3) to represent the transformed charge state of $n_z$ (n0, n1, n2, n3). 
\reva{Note that $w_z$ is an intermediate transfer variable used in the neural network to facilitate the training process, while $n_z$ represents the normalized charge state. Both $w_z$ and $n_z$ are presented in non-dimensional units in the plot. }

\begin{figure}[!htb]
\centering
    \includegraphics[width=0.6\linewidth]{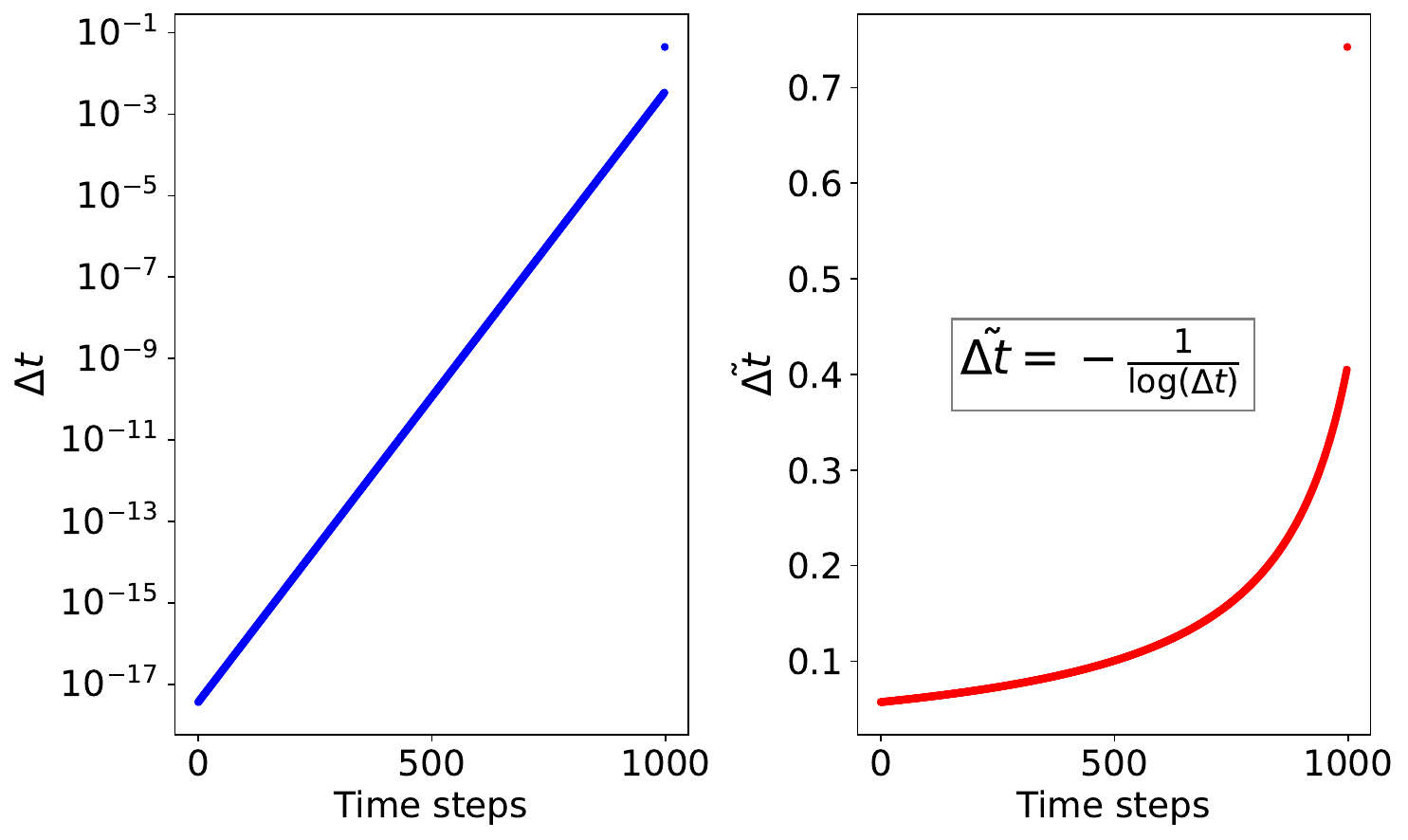}
    \caption{Scaling of the time steps.}
    \label{fig:time step_scaling}
\end{figure}

\paragraph{\textbf{Time Step Scaling.}} Time step size scaling is another crucial aspect for a successful training. For stiff equations or rapidly changing dynamics, a fixed time step size is often inadequate for capturing the system's behavior accurately. In such cases, adaptive time stepping methods, based on the local behavior of the system, are employed. This adaptive approach often results in more accurate and computationally efficient simulations. In the high-fidelity numerical simulation of the CR model, we used prescribed adaptive time steps in the numerical integration. 
{The dataset was collected from non-uniform time step solutions with $\Delta t_n$ ranging from $10^{-16}$ to $10^{-1}$ (unit second)}. The tiny scale of the time step size makes it challenging to train the neural network, as we input $\Delta t_n$ into the flow map to learn the latent dynamics evolution from current step $L(t_n$) to the next $L(t_{n+1})$. To make the neural network training more efficient, we use the following transfer formula to scale $\Delta t_n$ into a proper range of $(0, 1)$,
\begin{equation}
\tilde{\Delta t} = -\frac{1}{\log(\Delta t)}.
\end{equation}
See Fig.~\ref{fig:time step_scaling} for a demonstration.

\begin{figure}[!htb]
\centering
        \includegraphics[width=0.8\linewidth]{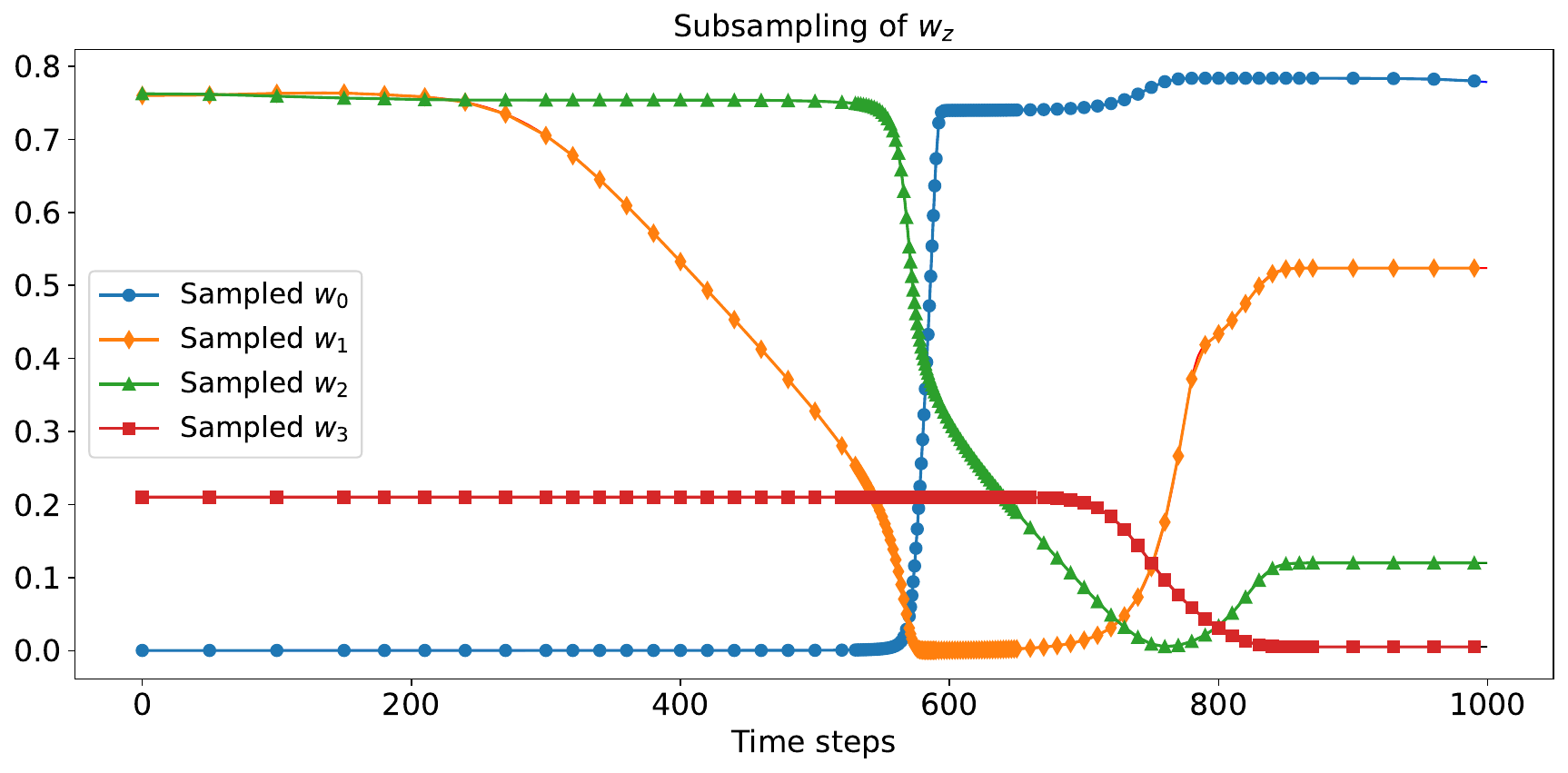}
    \caption{Resampled data points from the true trajectory.}
    \label{fig:coarse_sample}
\end{figure}

\paragraph{\textbf{Resampling.}} We simulate the CR equation~\eqref{eqn:rate-matrix} with 37 temperatures $T_e$ from 5eV to 95eV with a step size of 2.5, 10 different total densities \reva{with $n_A=$[1e14, 2e14, 3e14, 4e14, 5e14, 6e14, 7e14, 8e14, 9e14, 1e15] ($m^{-3}$), and 40 different initial conditions. Here we let the initial population for the states to be uniformly small except the ground states. Under different initial conditions, the final excited state accounts for different percentage of the total population $n_A$, ranging from 0.01\% to 2.2\%.} This results in a total of 14,800 trajectories. The dataset was collected with 1,000 time steps for each trajectory, leading to 14.8 million pairs of $(L_n, L_{n+1})$ in the flow map neural network training to learn latent dynamics. This large dataset is very expensive to train in practice. Additionally, there are a significant number of tiny time steps in each run, of which the corresponding flow maps are near-identity mappings (the first 500 time steps). This poses two challenges: first, the dataset size is substantial, and second, the presence of many near-identity mappings makes the training process difficult. To address these issues, we performed coarse sampling of the data. The purposes of this approach were to reduce the dataset size and to avoid the identity mappings. Instead of using all 1,000 time step data points, we performed coarse sampling to retain 161 points; see Fig.~\ref{fig:coarse_sample}.
Note that time steps are selected to fully resolve the sharp transition of the dynamics.


\subsection{\label{sec:AE}Autoencoder Training}
\begin{figure}[!htb]
\centering
    \includegraphics[width=0.6\linewidth]{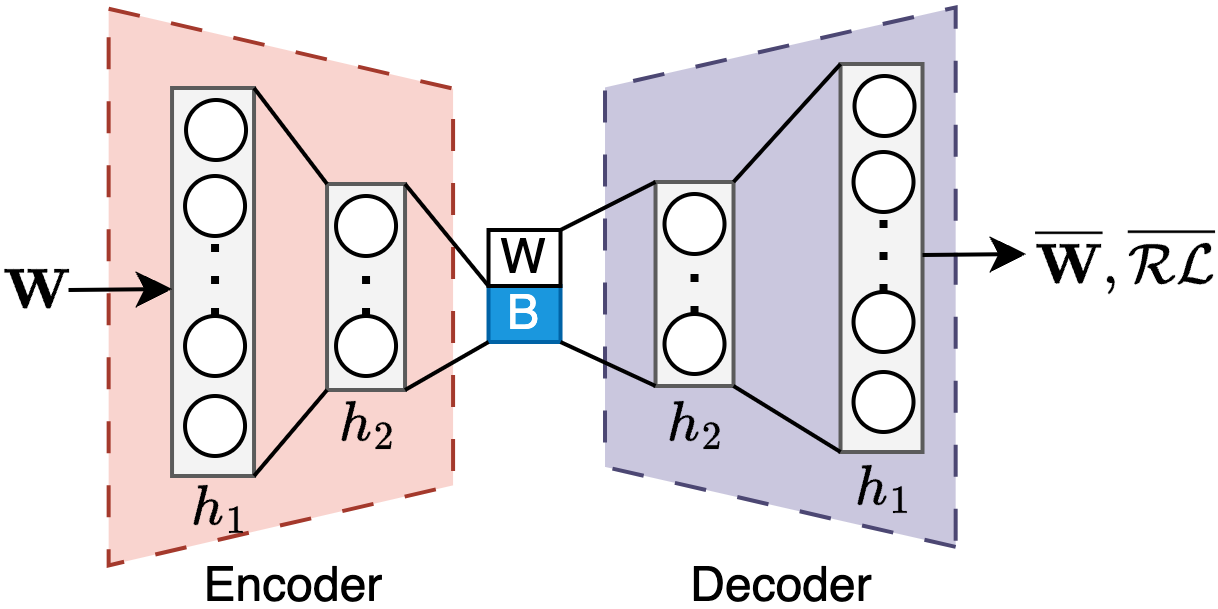}
    \caption{The autoencoder architecture, $h_1,h_2$ are the hidden layers with units 64 and 32 respectively}
    \label{fig:AE}
\end{figure}

\begin{figure}[!htb]
\centering
    \includegraphics[width=1\linewidth]{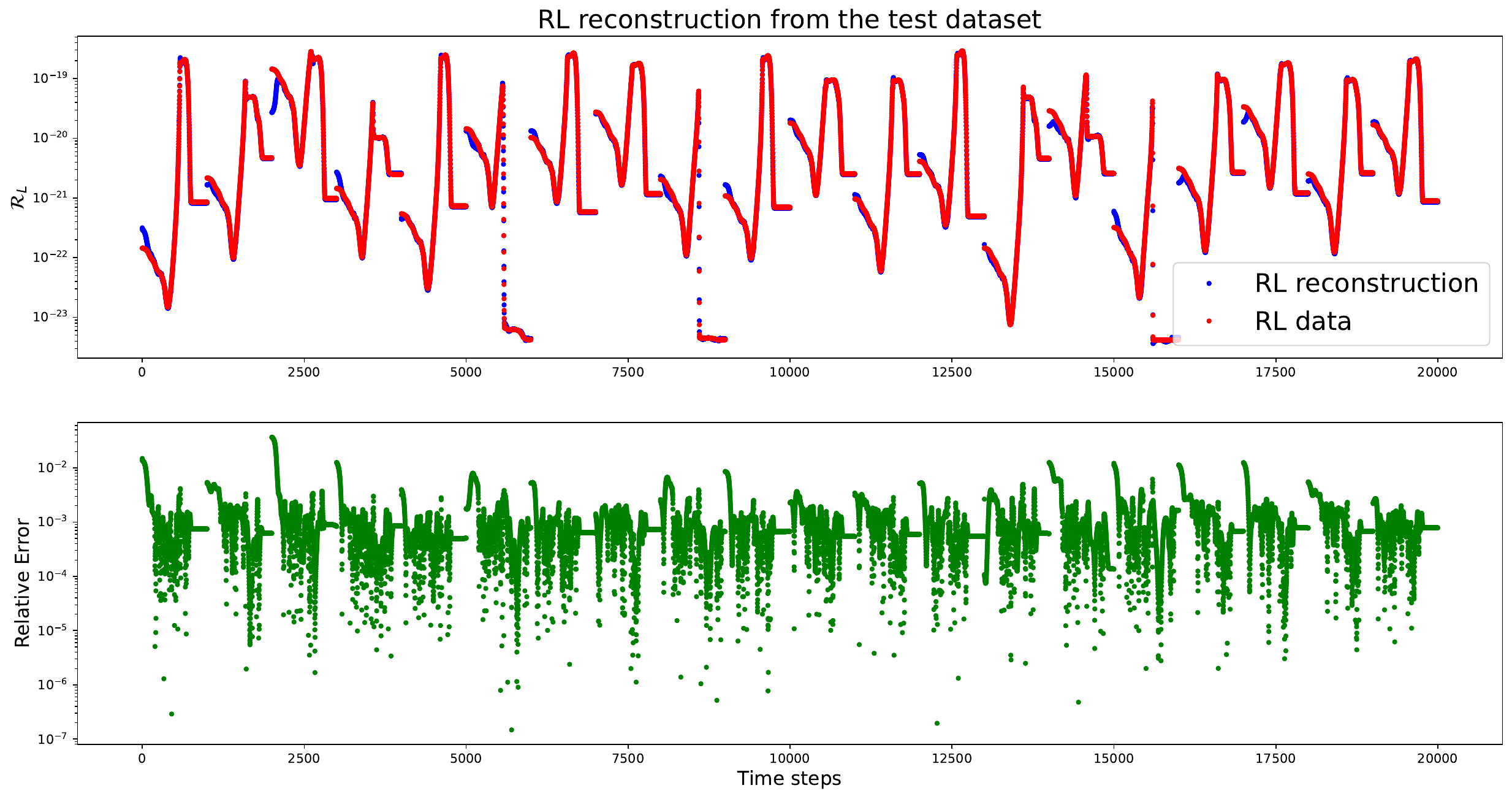}
    \caption{\reva{Top: The radiative loss rate $\mathcal{R}_L$ reconstruction from the autoencoder neural network. 20 randomly sampled trajectories from the test dataset are plotted. Red points are the true data, blue dots are the prediction from the decoder. All trajectories share the same time steps (1000 time steps), and each trajectory corresponds to a separate simulation for the respective parameter. Bottom: The relative error over all the time steps for each trajectory tested in the top figure.}}
    \label{fig:AE_RL}
\end{figure}

The entire model is trained in two steps. First, we train the autoencoder to identify the latent space. Next, we use a flow map neural network (FMNet) to learn the latent dynamics. In the autoencoder architecture, the encoder and decoder both have two hidden layers. The input is the normalized ion population density $\mathbf{W}$ with an input dimension of 94. The output of the decoder is the reconstruction of the ion population density and radiative loss rate $\mathcal{R}_L$, with a dimension of 95. Increasing the number of hidden neurons and the latent space size increases the training cost. For the fully connected layers, sigmoid activations are used. The learning rate has constant decay after every 1000 epochs starting with $0.001$.  We trained the model for 10,000 epochs, experimenting with different numbers of hidden units and latent space dimensions. To balance accuracy and efficiency, using $h1=64, \, h2=32$, and a latent space of dimension 10 
(it consists of a black space of dimension 6 and the white space of dimension 4) provides a reasonably accurate reconstruction and prediction of radiative loss. 
Fig.~\ref{fig:AE} shows the autoencoder architecture. \reva{The autoencoder neural network is trained on 70\% of the total trajectory data and evaluated on a test dataset containing 600 trajectories with different combination of initial conditions, temperature ($T_e$), and density ($n_A$)). Fig.~\ref{fig:AE_RL} shows the reconstruction of $\mathcal{R}_L$ from 20 randomly sampled testing trajectories data (data unseen by the neural network). } The model is trained using one NVIDIA A100 GPU. 
Under this configuration, the training cost is 27 hours. We used this latent space configuration for our flow map training. 

\subsection{Flow Map Training}

\subsubsection{\label{sec:epred}Prediction Error}
\begin{figure}[htb]
\centering
        \includegraphics[width=0.9\linewidth]{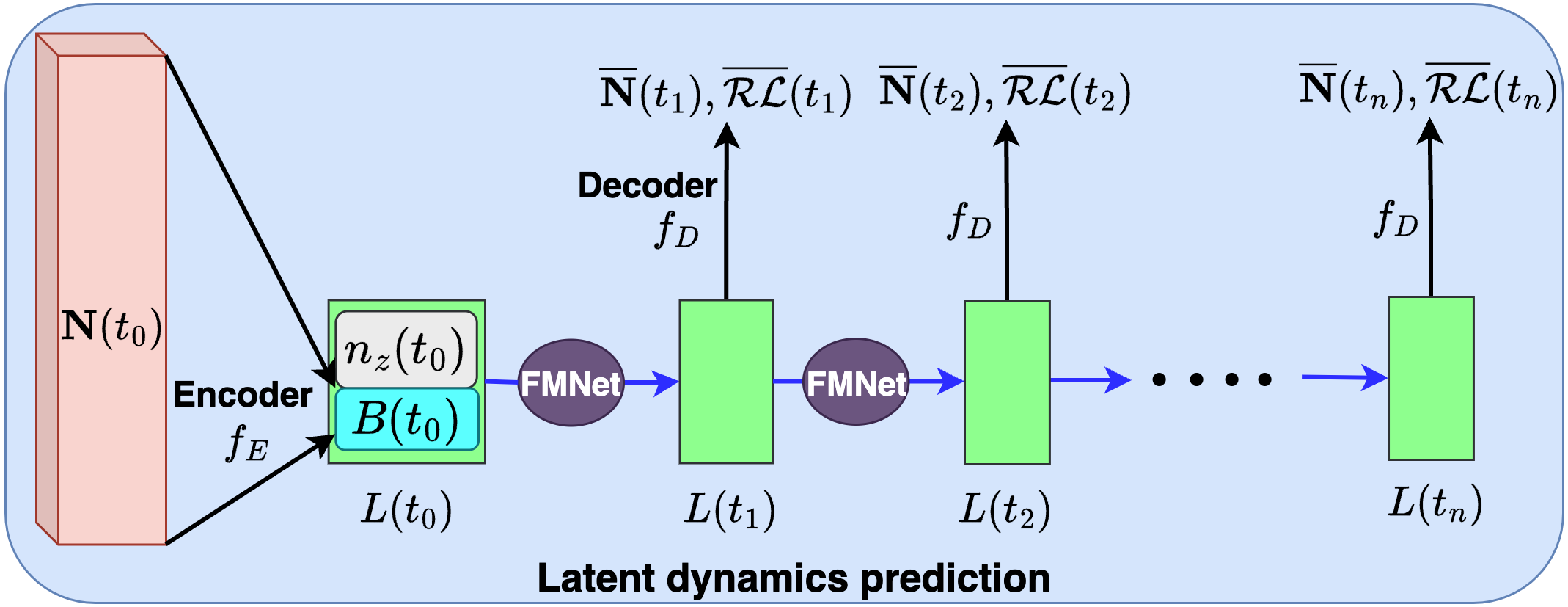}
    \caption{The iterative prediction scheme of the latent dynamics and radiative loss rate $\mathcal{R}_L$ after the autoencoder and flow map network (FMNet) are trained.}
    \label{fig:traj_pred_demo}
\end{figure}

In the prediction phase, we provide the initial condition (latent variable at $t_0$ projected by the encoder) to the FMNet. FMNet then iteratively predicts the latent trajectory, which includes both the charge state dynamics (white space) and the unknown dynamics (black space); see Fig.~\ref{fig:traj_pred_demo}. This latent trajectory is subsequently fed into the decoder to obtain the radiative loss rate $\mathcal{R}_L$. 
The prediction error of the charge state, used in our model evaluation, is defined as the Mean Squared Error (MSE) at each time step,
\begin{equation*}
e_{pred} = \frac{1}{S}\sum_{i=1}^S \|n^i_z-\overline{n^i_z}\|^2
\end{equation*}
It is important to note that the training error is computed based on one-step predictions, while the prediction error in the testing phase accumulates at every time step, reflecting the compound effect of iterative predictions. Using prediction errors to evaluate model performance ensures the robustness and reliability of the model, making it suitable for practical applications in predicting radiative loss rates and understanding charge state dynamics. By leveraging the latent space representation, the reduced-order model effectively reduces computational complexity while maintaining high accuracy. This approach facilitates efficient and accurate simulations in high-fidelity numerical experiments, thereby enhancing the model's utility in real-world scenarios.

\subsubsection{Dynamics Prediction from Different Initial Conditions}
\begin{figure}[!htb]
\centering
        \includegraphics[width=0.9\linewidth]{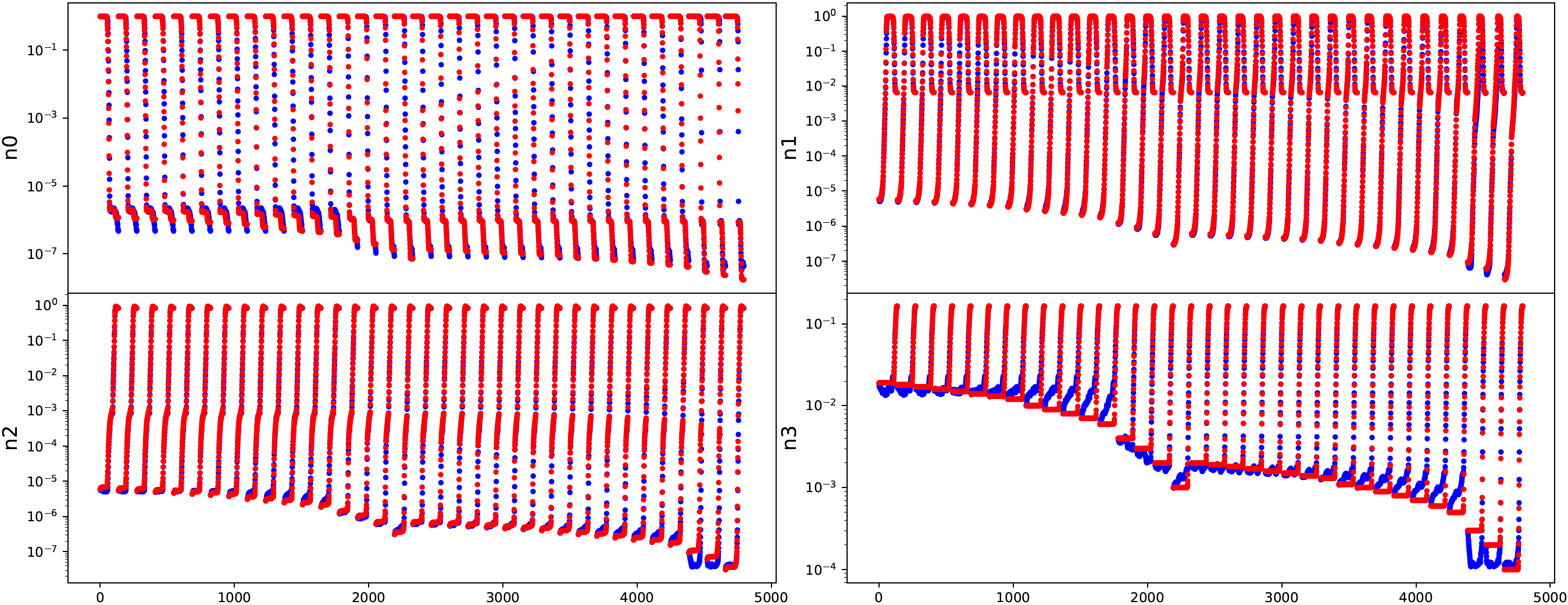}
    \caption{Charge state $n_z$ trajectory prediction from the flow map neural network (FMNet). Each trajectory represent different initial conditions. Red dots are the true data, and blue dots are the FMNet prediction}
    \label{fig:40traj_prednz}
\end{figure}

\begin{figure}[!htb]
\centering
        \includegraphics[width=0.9\linewidth]{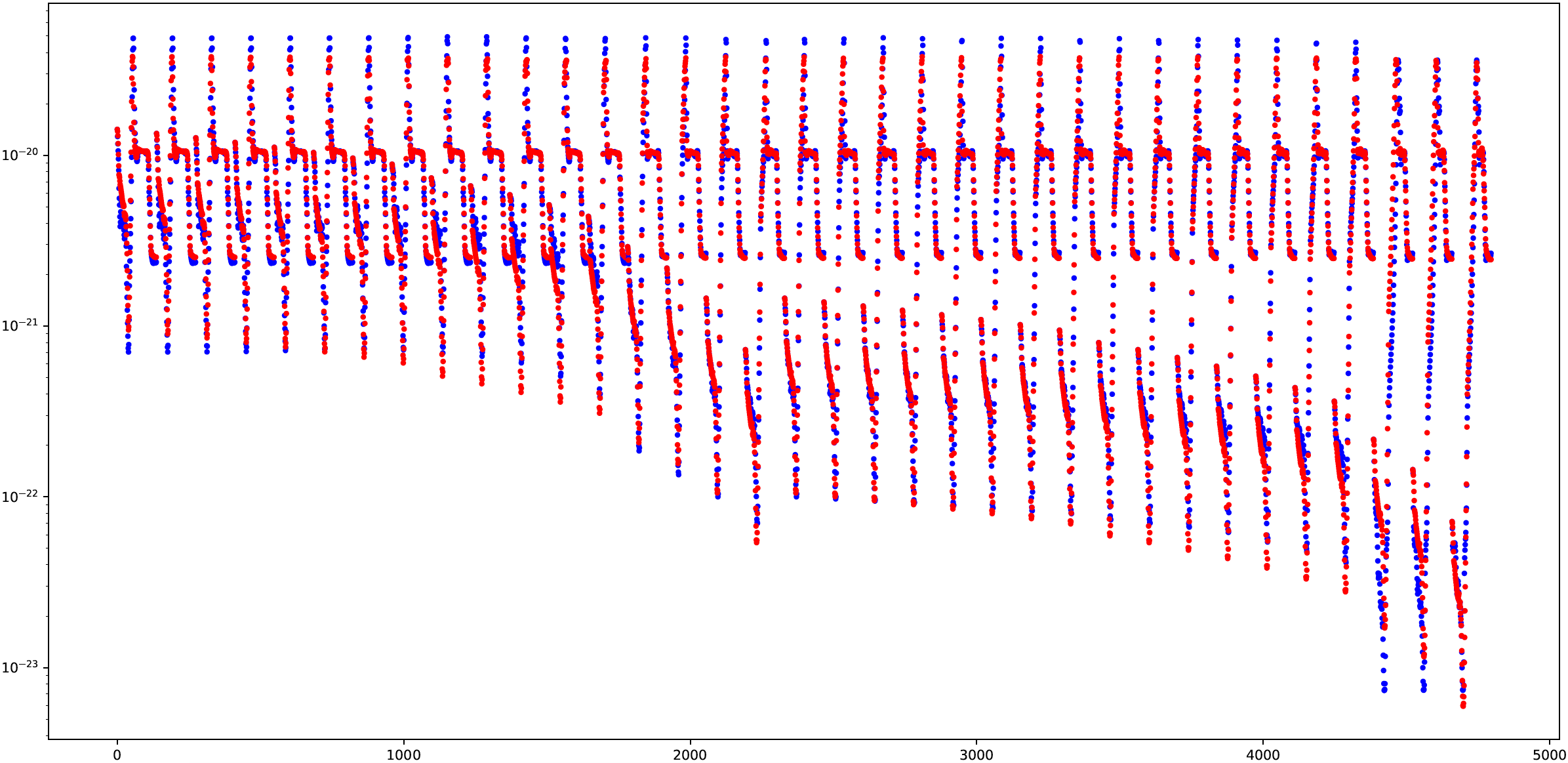}
    \caption{Radiative loss rate prediction from the decoder after feeding the predicted latent dynamics.}
    \label{fig:40traj_predRL}
\end{figure}

\begin{figure}[!htb]
\centering
        \includegraphics[width=0.9\linewidth]{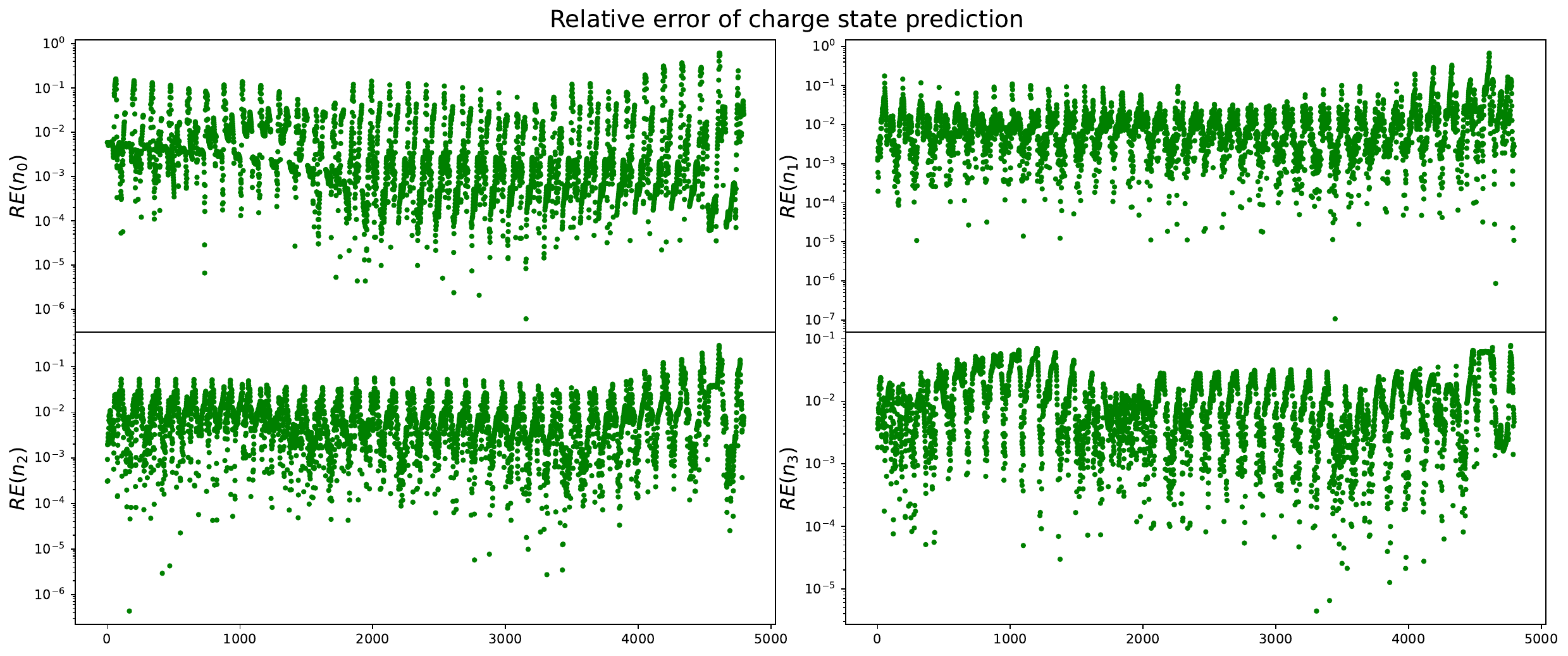}
    \caption{{Relative error of charge state $n_z$ trajectory prediction in Fig.~\ref{fig:40traj_prednz}}}
    \label{fig:40traj_relaer_nz}
\end{figure}

\begin{figure}[!htb]
\centering
        \includegraphics[width=0.9\linewidth]{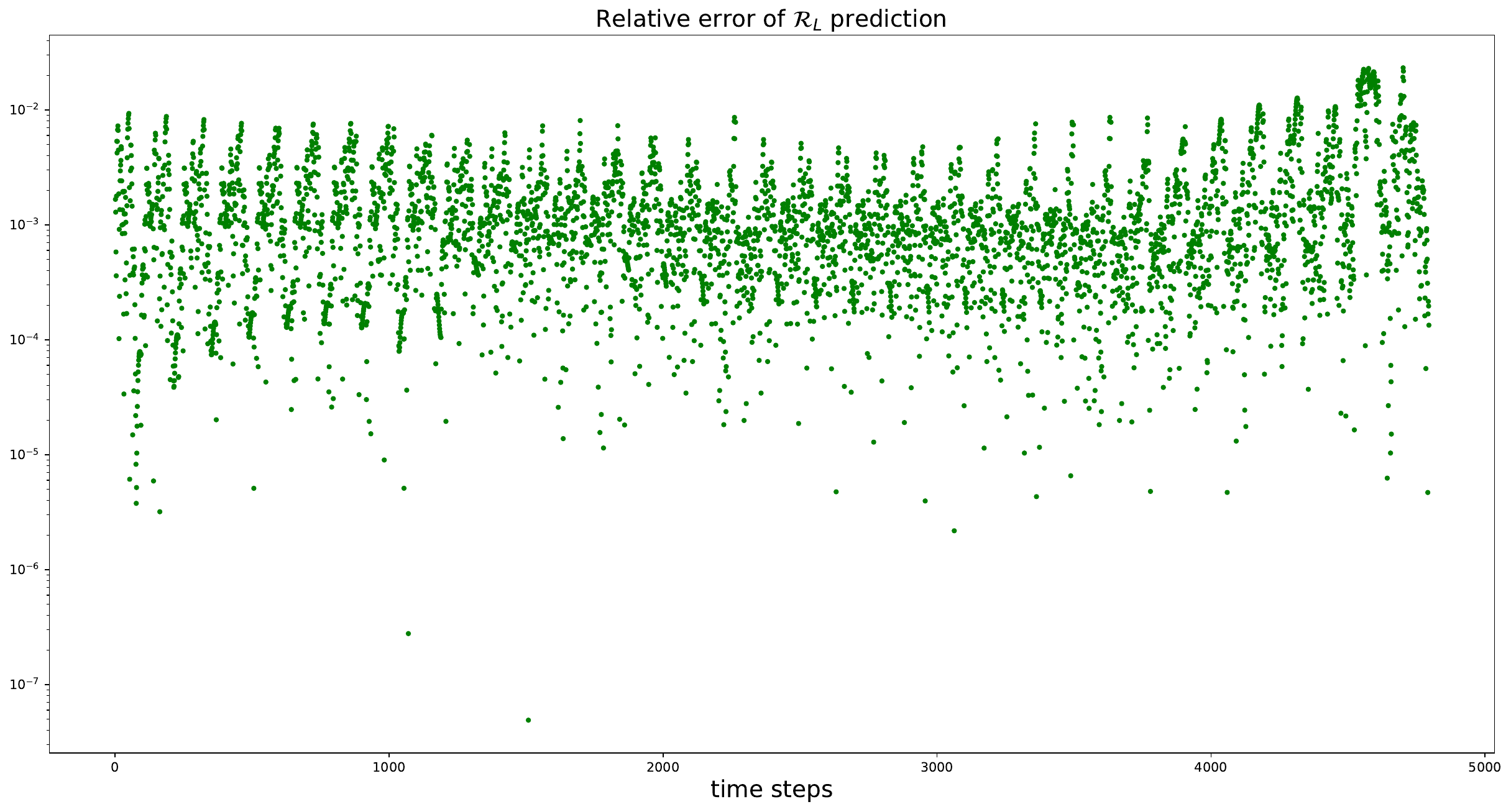}
    \caption{{Relative error of radiative loss rate $\mathcal{R}_L$ in Fig.~\ref{fig:40traj_predRL}}}
    \label{fig:40traj_relaer_RL}
\end{figure}

We first evaluate the performance of the FMNet under various initial conditions. We initialize the FMNet with different latent variables, each corresponding to a unique initial condition in the high-fidelity simulation data. The purpose of this evaluation is to assess the model's ability to generalize and accurately predict the CR dynamics from different starting points. For each initial condition, the FMNet iteratively predicts the latent trajectory, capturing the evolution of charge state dynamics and unknown dynamics. The predicted latent trajectories are then decoded to obtain the corresponding radiative loss rates $\mathcal{R}_L$.
We quantify the prediction accuracy using metrics such as Mean Squared Error (MSE) and Mean Absolute Error (MAE) across all time steps for each initial condition. The results are compared to the ground truth obtained from the high-fidelity simulations. {For this test, the density is fixed at $n_A=1e14$ ($m^{-3}$) and temperature is set to $T_e=35(eV)$. We set different initial conditions by varying the final excited state as a percentage of the total population $n_A$, ranging from 0.01\% to 2.2\%. In the numerical test, we uniformly sampled 40 values between 0.01\% and 2.2\% to generate the initial conditions. Fig.~\ref{fig:40traj_prednz} illustrates the prediction performance for a subset of initial conditions, showing both the predicted and true trajectories of key variables. Fig.~\ref{fig:40traj_predRL} shows the corresponding radiative loss predictions. Figs.~\ref{fig:40traj_relaer_nz} and~\ref{fig:40traj_relaer_RL} are the corresponding relative error at each time step.}
Our result indicates that the model maintains robust performance across a wide range of initial conditions, with prediction errors remaining within acceptable bounds. This demonstrates the model's capability to adapt to different starting points and accurately capture its dynamics.

\subsubsection{Dynamics Prediction from Different Parameters}
\begin{figure}[htp]
    \centering
    \includegraphics[width=0.6\linewidth]{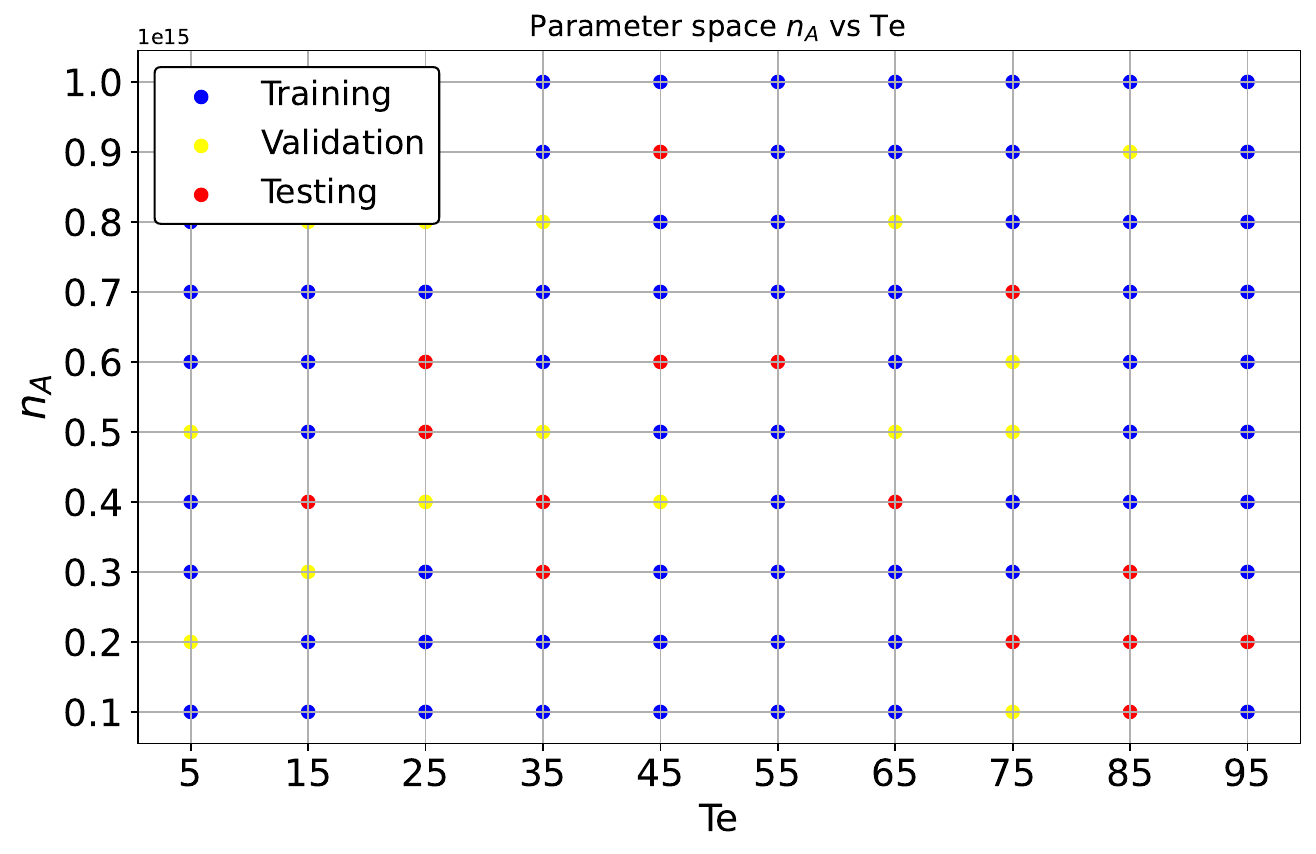}
    \caption{Parameters sampled for training and prediction test.}
    \label{fig:param_split}
\end{figure}

\begin{figure}[htp]
    \centering
    \includegraphics[width=1.0\linewidth]{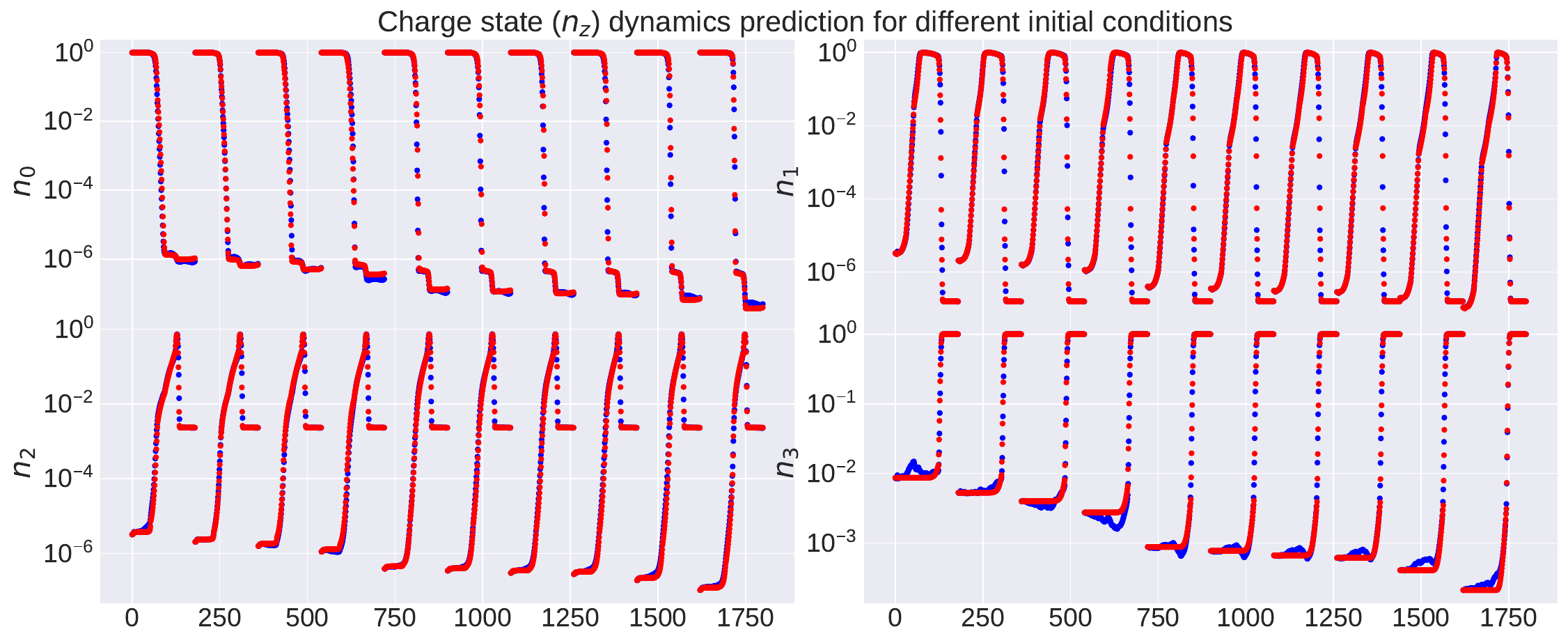}
    \caption{Charge state trajectory prediction at temperature $T_e=65$, $n_A=$5e14 with different initial conditions in the testing dataset, 10 trajectories are plotted. Red represents true data, and blue represents the model prediction.}
    \label{fig:pred_Te65_nz}
\end{figure}

\begin{figure}[htp]
    \centering
    \includegraphics[width=1.0\linewidth]{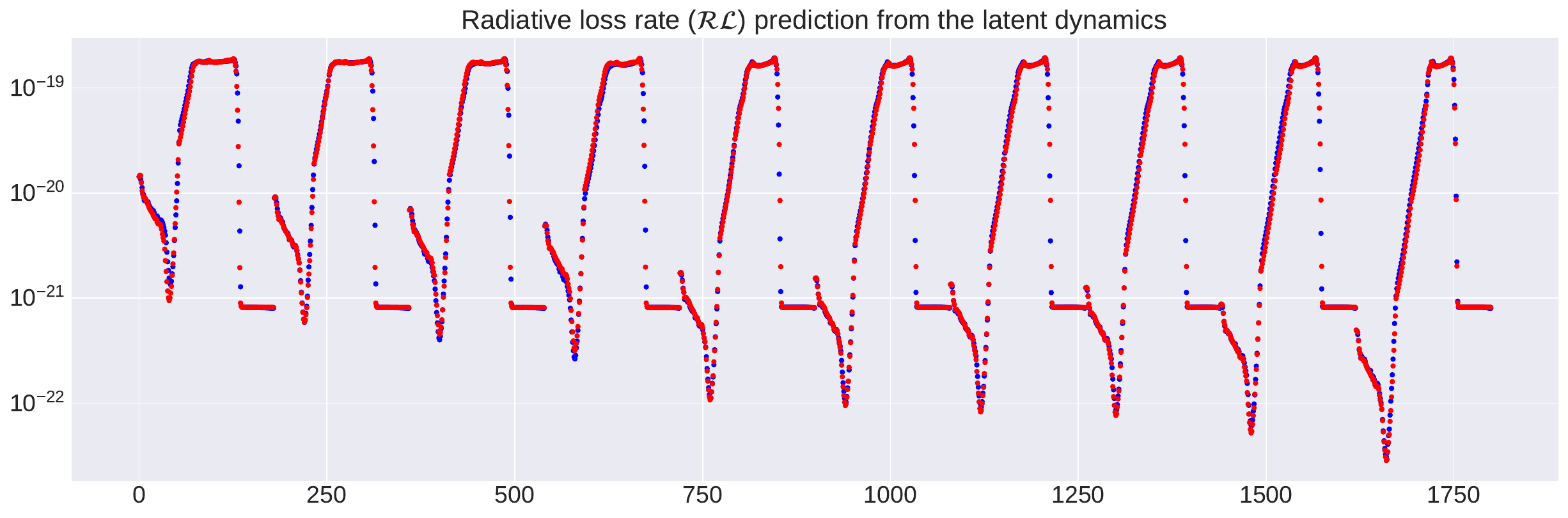}
    \caption{Radiative loss rate prediction at temperature $T_e=65$, $n_A=$5e14 with different initial conditions in the testing dataset. Red represents true data, and blue represents the model prediction.}
    \label{fig:pred_Te65_RL}
\end{figure}

\begin{figure}[htp]
    \centering
    \includegraphics[width=1.0\linewidth]{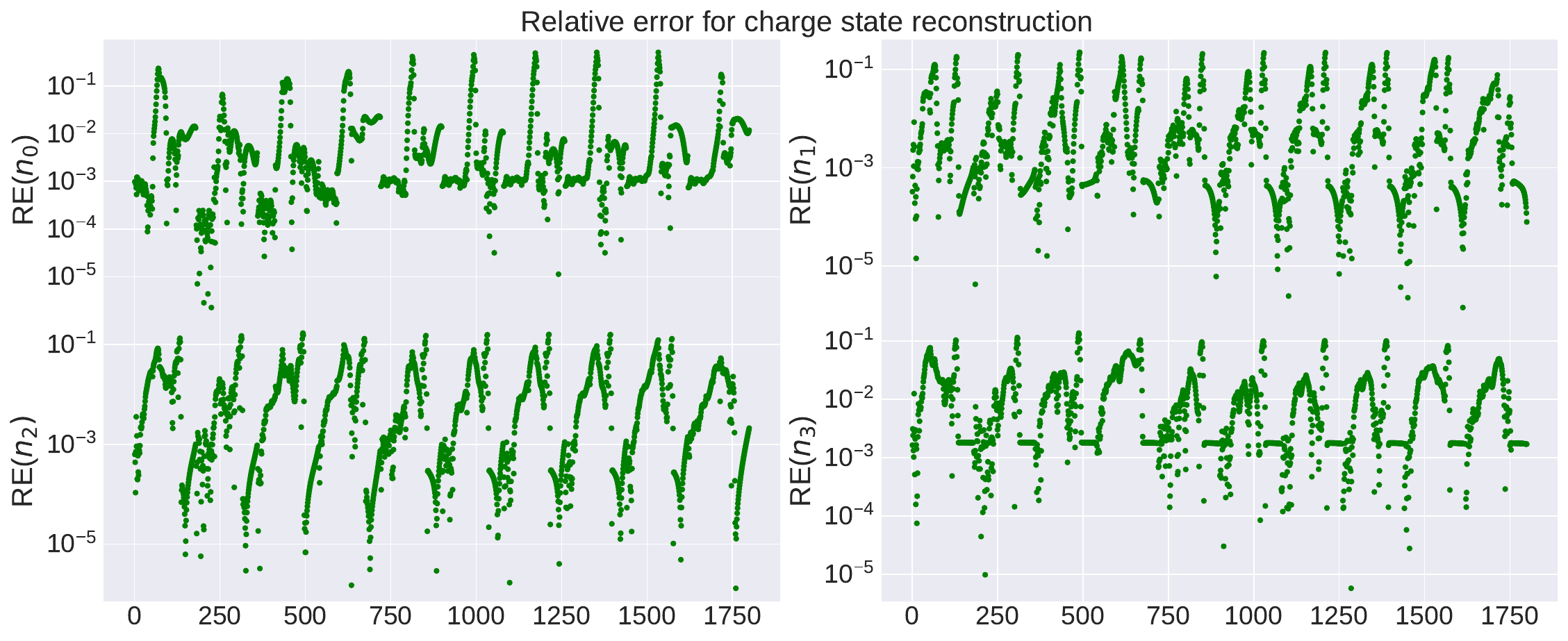}
    \caption{\reva{Relative error (RE) of charge state trajectory prediction at temperature $T_e=65$ in Fig.~\ref{fig:pred_Te65_nz}}}
    \label{fig:relaer_te65_nz}
\end{figure}

\begin{figure}[htp]
    \centering
    \includegraphics[width=1.0\linewidth]{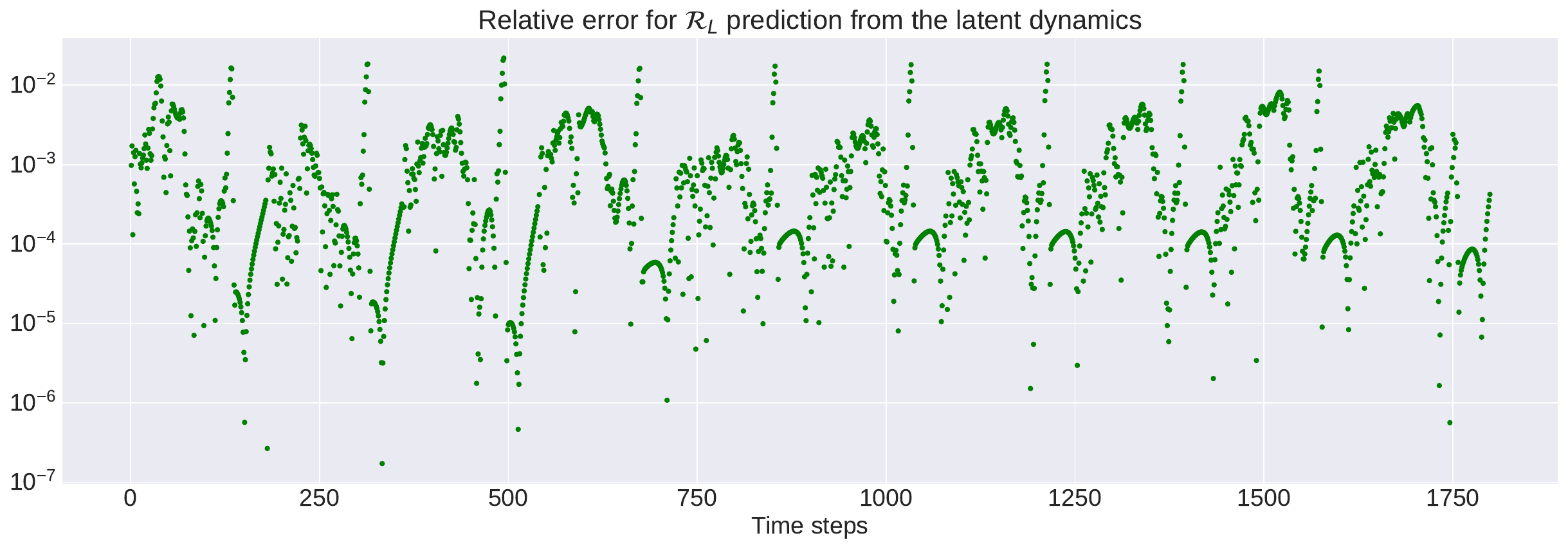}
    \caption{\reva{Relative error (RE) of radiative loss rate prediction at temperature $T_e=65$ in Fig.~\ref{fig:pred_Te65_RL}}}
    \label{fig:relaer_te65_rl}
\end{figure}

\begin{figure}[htp]
    \centering
    \includegraphics[width=1.0\linewidth]{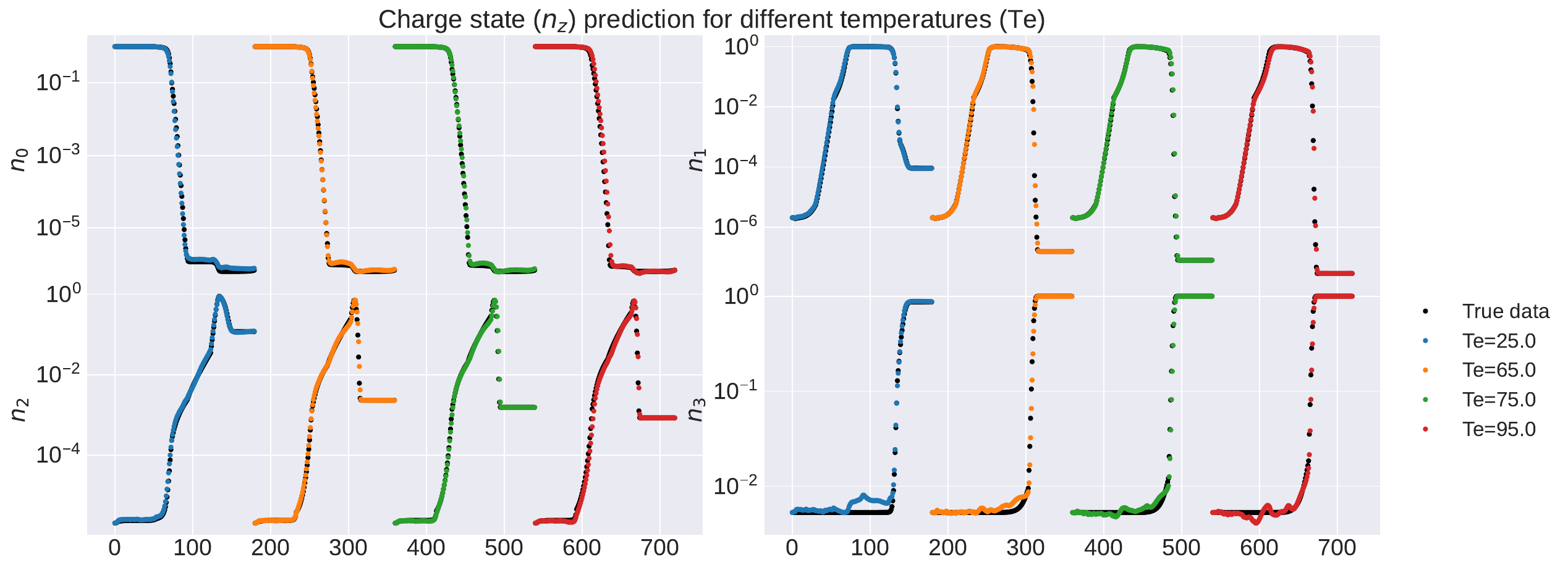}
    \caption{Charge state trajectory prediction for different temperatures $T_e$ in the testing dataset. Dark represents true data, and 4 different colors represent the model prediction from different temperatures. }
    \label{fig:pred_Tetest_nz}
\end{figure}

\begin{figure}[htp]
    \centering
    \includegraphics[width=0.9\linewidth]{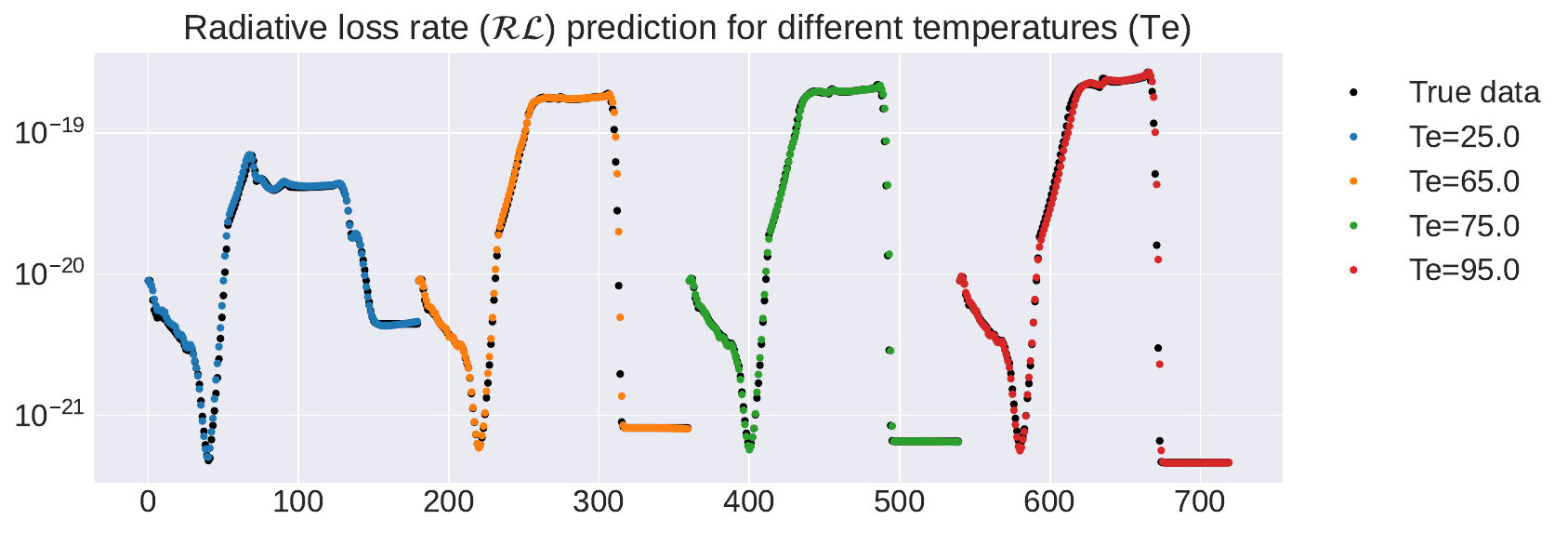}
    \caption{Radiative loss rate prediction for different temperatures $T_e$ in the testing dataset. Dark represents true data, and 4 different colors represent the model prediction from different temperatures. }
    \label{fig:pred_Tetest_RL}
\end{figure}

\begin{figure}[htp]
    \centering
    \includegraphics[width=1.0\linewidth]{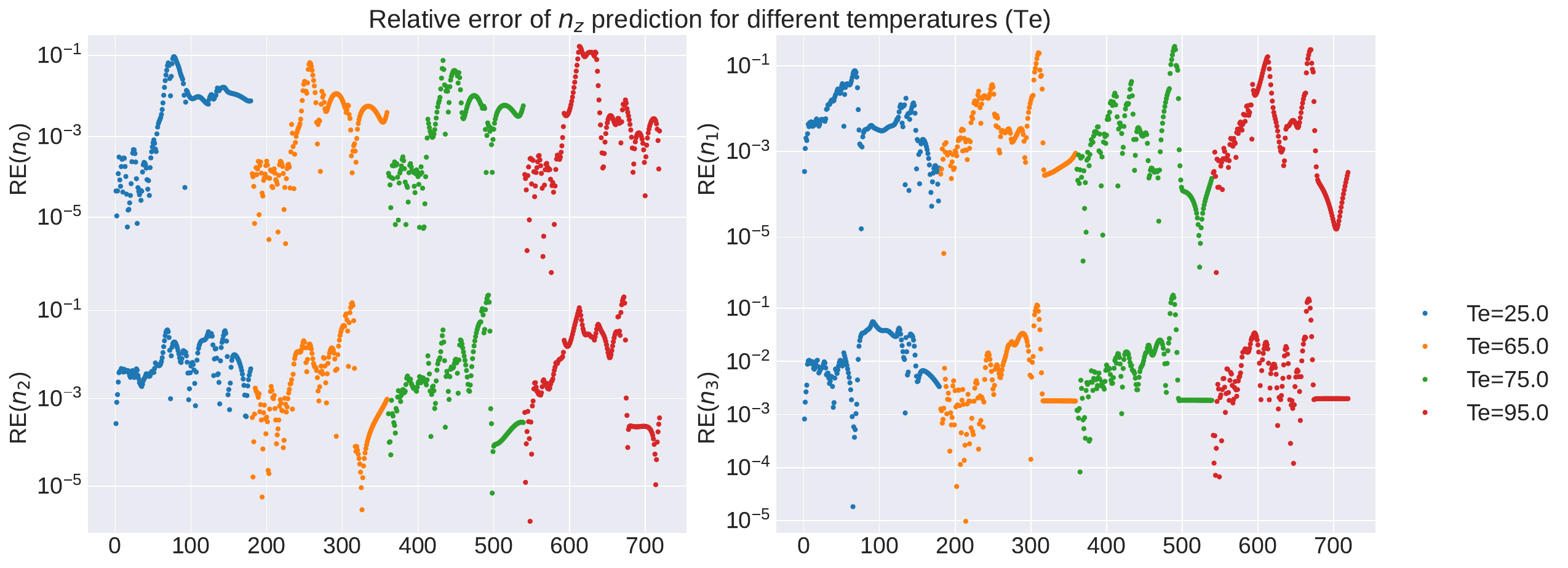}
    \caption{\reva{Relative error (RE) of charge state trajectory prediction for different temperatures $T_e$ in Fig.~\ref{fig:pred_Tetest_nz} }}
    \label{fig:relaer_Tewz}
\end{figure}

\begin{figure}[htp]
    \centering
    \includegraphics[width=0.9\linewidth]{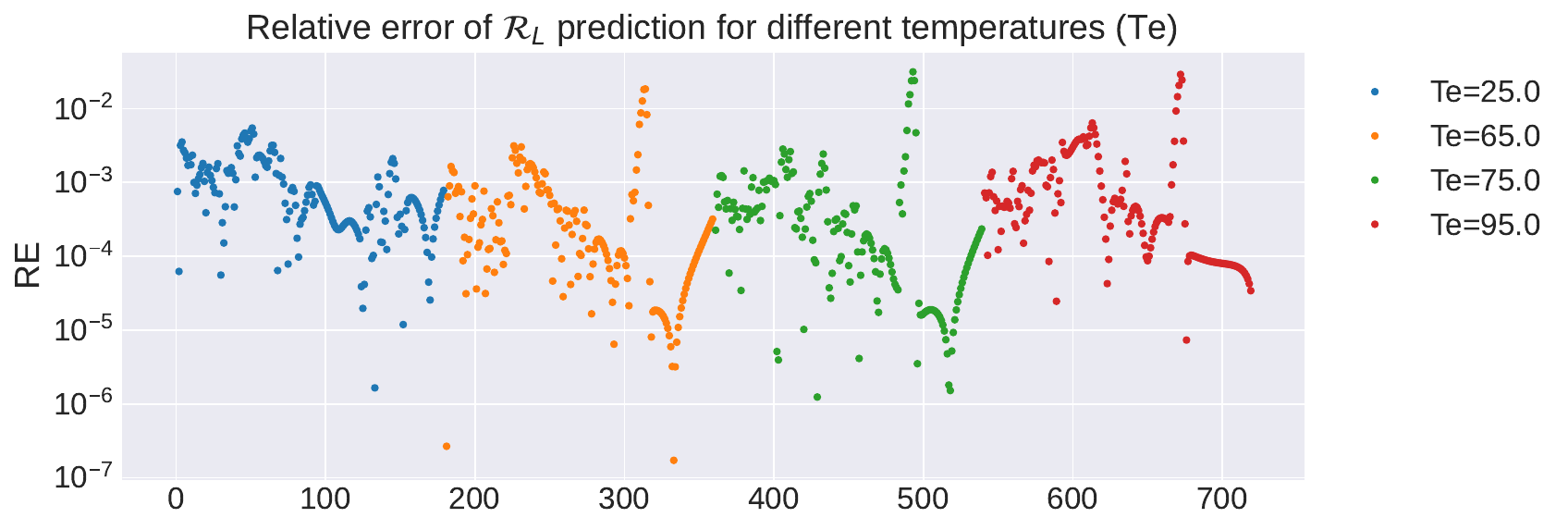}
    \caption{\reva{Relative error (RE) of radiative loss rate prediction for different temperatures $T_e$ in Fig.~\ref{fig:pred_Tetest_RL}}}
    \label{fig:relaer_Terl}
\end{figure}

In this section, we extend the evaluation to different parameter settings for total density $n_A$ and electron temperature $T_e$. These parameters play a crucial role in the behavior of the CR model, influencing the rates of collisional and radiative processes. To assess the model's performance across different parameter values, we generate predictions for various combinations of $n_A$ and $T_e$. The FMNet is trained to account for these parameters as inputs, enabling it to adapt its predictions based on the specific conditions.
We systematically vary $n_A$ and $T_e$  within their respective ranges used in the high-fidelity simulations, Fig.~\ref{fig:param_split} shows the split of the parameters in dataset for training, validation and testing. For each combination of $n_A$ and $T_e$, the model predicts the latent trajectory and the corresponding radiative loss. We then compare these predictions to the ground truth data, using prediction error. {Figs.~\ref{fig:pred_Te65_nz} and \ref{fig:pred_Te65_RL} show the prediction results of charge state and radiative loss from the testing dataset at $T_e=65$ (eV) and $n_A=5e14$ ($m^{-3}$). Figs.~\ref{fig:relaer_te65_nz} and~\ref{fig:relaer_te65_rl} plot the corresponding relative error over each time step. Figs.~\ref{fig:pred_Tetest_nz} and~\ref{fig:pred_Tetest_RL} plot the prediction for different values of $T_e$ ; Figs.~\ref{fig:relaer_Tewz} and~\ref{fig:relaer_Terl} are the corresponding relative error over time. These plots illustrate the model's ability to accurately capture the dynamics under varying conditions.
Our findings suggest that the model performs well across a broad spectrum of parameter values, maintaining high accuracy in its predictions.} This highlights the model's flexibility and robustness, making it a valuable tool for simulating and understanding the behavior of the CR system under different physical conditions.

\subsubsection{Neural Network Architecture Search}
Neural network architecture search (NAS) is a crucial process in the development of ML models, focusing on automating the design of optimal neural network architectures. The idea behind NAS is to systematically explore a vast search space of possible architectures to identify the most effective configurations that meet specific performance criteria, such as accuracy, efficiency, and computational cost.

\begin{figure}[!htb]
        \includegraphics[width=1.0\linewidth]{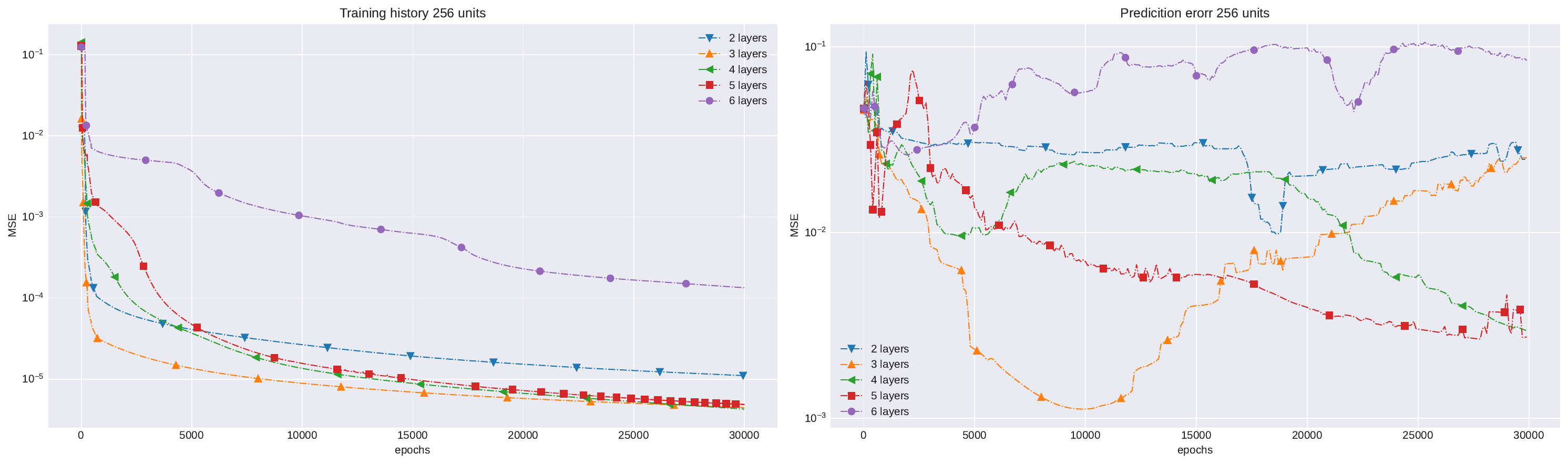}
    \caption{Training history of the FMNet. Training error (left) vs prediction error (right) from different network architectures.}
    \label{fig:fmnet_hist}
\end{figure}

\begin{figure}[!htb]
        \includegraphics[width=0.6\linewidth]{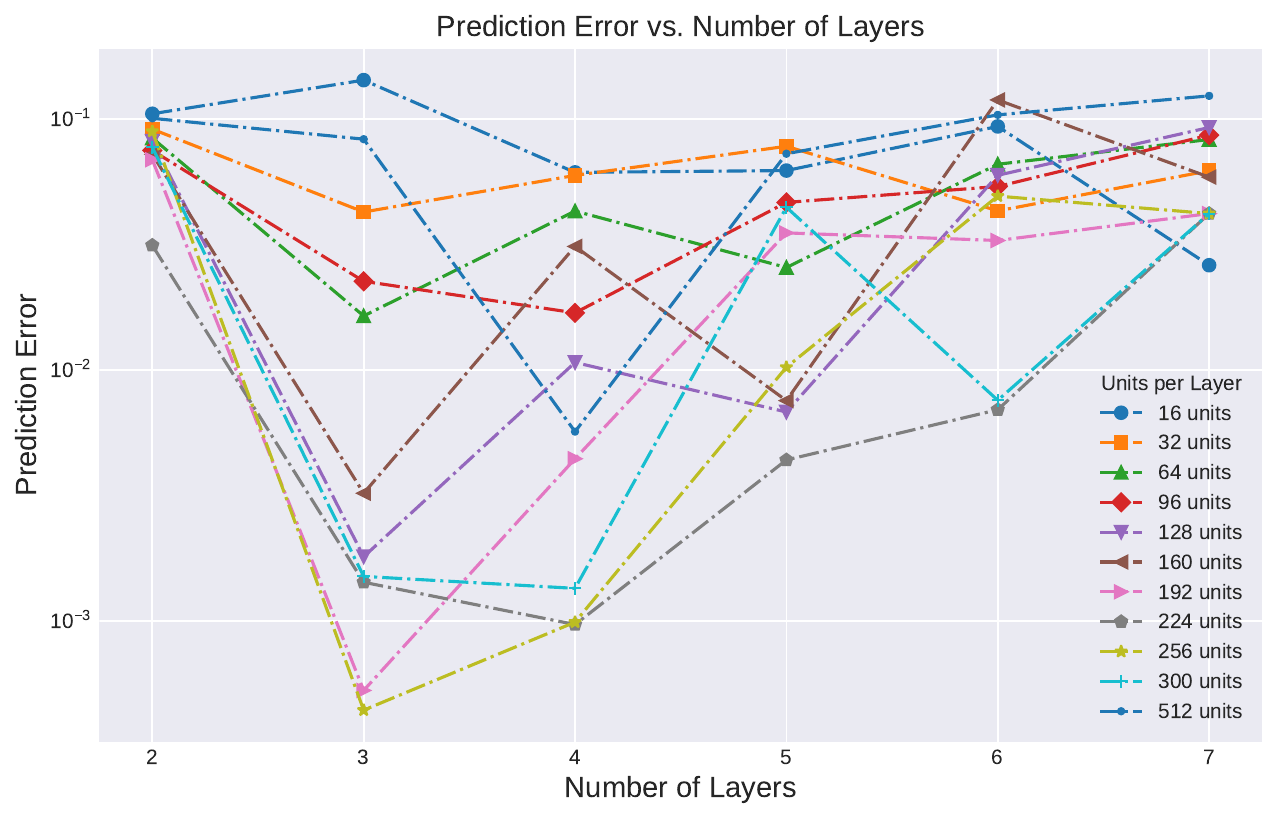}
    \caption{\reva{Prediction error from models with different architecture. The FMNet reaches its best prediction error with 3 layers and 256 units for each layer.  }}
    \label{fig:2derrvsunits}
\end{figure}

In practice, NAS involves defining a search space that specifies the range of possible architectures, including the number of layers, the type of layers (e.g., convolutional, fully connected), and the number of units in each layer. The search algorithm then navigates this space to find architectures that maximize a given performance metric on a validation dataset. In our work, we employed a grid search methodology to systematically explore a range of possible configurations. The primary objective was to determine the optimal architecture by varying the number of layers and the number of hidden units within each layer. The number of layers range set was from 2 to 7 layers. This range includes both simpler models with fewer layers, which may train faster and are less prone to overfitting, and more complex models with additional layers, which have the capacity to capture more intricate patterns in the data. For each layer, the number of hidden units was varied between 16 and 512. By systematically combining these two parameters (number of layers and hidden units), the grid search examined a wide array of architectures and it gives an initial study on the optimal neural network structure for our training data. \reva{The results show that the FMNet nearly reaches its best prediction error with 3 layers and 256 units for each layer; see Fig.~\ref{fig:2derrvsunits}. The behavior observed in Fig.~\ref{fig:fmnet_hist}, where the prediction error does not strictly decrease beyond 10,000 epochs despite the training error continuing to decrease, is related to the nature of the error accumulation in time-series predictions. The training error is based on a one-step prediction scheme, meaning that during training, the network is optimized to predict the next timestep given the current input. In this case, minimizing the training error over more epochs leads to better one-step predictions, which is why the training error keeps decreasing. However, the prediction error defined in~\ref{sec:epred} shown in Fig.~\ref{fig:fmnet_hist} corresponds to the network's performance over multiple timesteps. In this case, the error accumulates over time, as small errors in early predictions can propagate and compound over the entire trajectory. As a result, even though the network's one-step predictions improve with more training (i.e., lower training error), the cumulative prediction error over multiple timesteps may not decrease as consistently. This can explain why, after a certain number of epochs (around 10,000), the network's overall prediction accuracy begins to degrade, as the model may start overfitting to the training data, leading to less generalization in long-term predictions. In the training process, we employed an early stopping criterion to mitigate overfitting and preserve the neural network's generalization performance. Specifically, we monitored the prediction error on a testing dataset that was unseen by the neural network during training. When the prediction error on the testing dataset ceased to improve for a certain number of epochs, we saved the model's weights corresponding to the lowest testing error. This approach ensures that we prevent the model from overfitting to the training data, where the training error might continue to decrease, but the model's ability to generalize on unseen data could degrade. By using early stopping, we effectively capture the point where the model performs best on the testing data, thus preserving its predictive performance over unseen timesteps and preventing any further degradation in long-term predictions.}

Addressing NAS effectively requires balancing the exploration of diverse architectures with the exploitation of promising configurations. Techniques such as early stopping, weight sharing, and transfer learning are often employed to reduce the computational burden and accelerate the search process. As NAS continues to evolve, it holds the potential to significantly advance the field of neural network design, making it more accessible and efficient. The grid search we used in this study is computationally expensive and may miss optimal configurations lying between grid points. In the future work, we will explore Bayesian optimization and reinforcement learning for more robust search.

\subsubsection{Impact of Training Data Size}
The performance of neural networks is influenced by two key factors: the amount of training data available and the complexity of the model architecture. In the previous section, we used a grid search method to find the optimal neural network architecture. In this section, we explore how increasing the size of the training dataset and the complexity of the neural network model impacts prediction performance.

\begin{figure}[!htb]
\centering
        \includegraphics[width=1.0\linewidth]{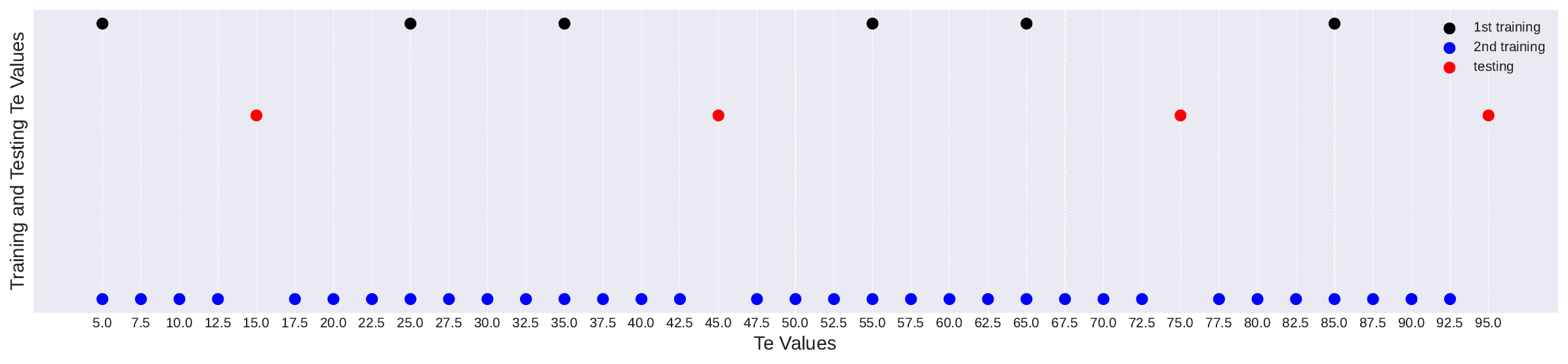}
    \caption{Data used for different training test}
    \label{fig:2DTe_values_test}
\end{figure}

\begin{figure}[!htb]
\centering
        \includegraphics[width=1.0\linewidth]{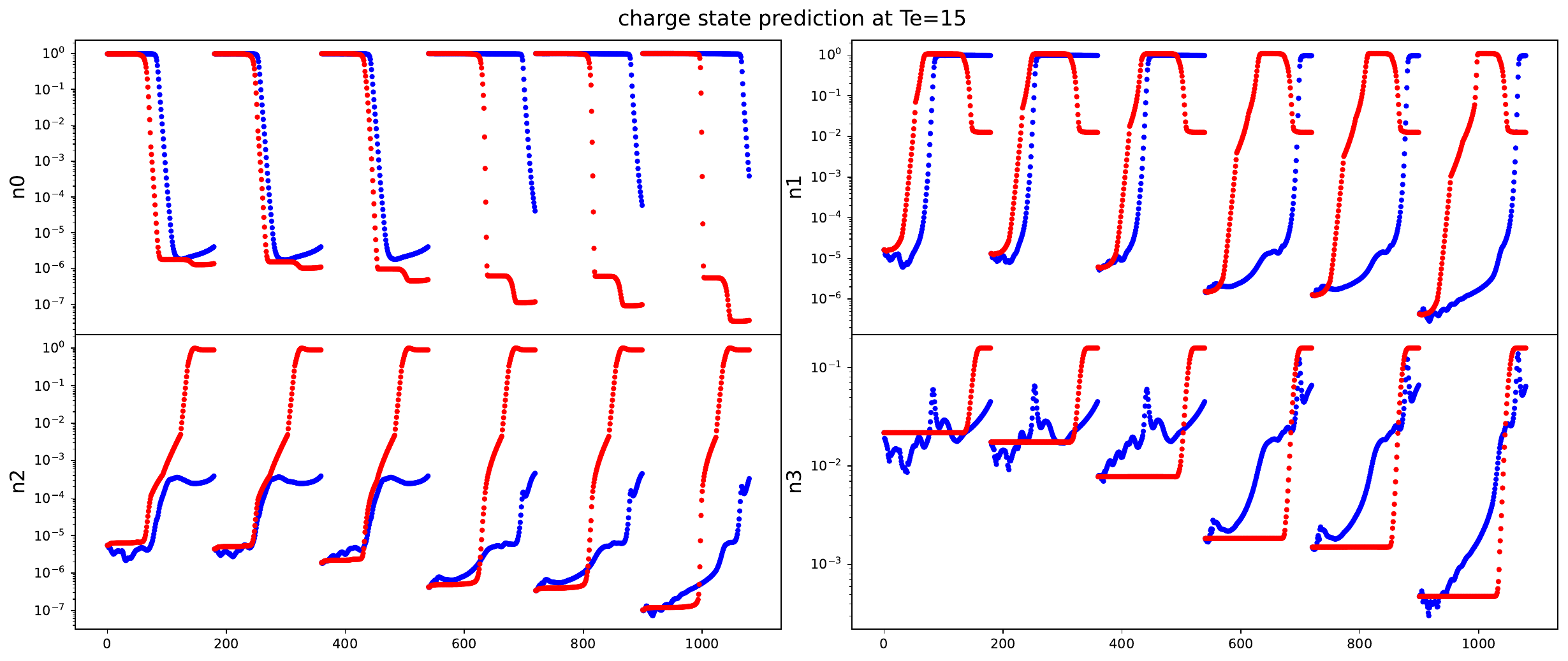}
    \caption{Charge state prediction from the FMNet trained with first training $T_e$ data (see dark dots in Fig.~\ref{fig:2DTe_values_test}). }
    \label{fig:pred_1stdata}
 \end{figure}

\begin{figure}[!htb]
\centering
        \includegraphics[width=1.0\linewidth]{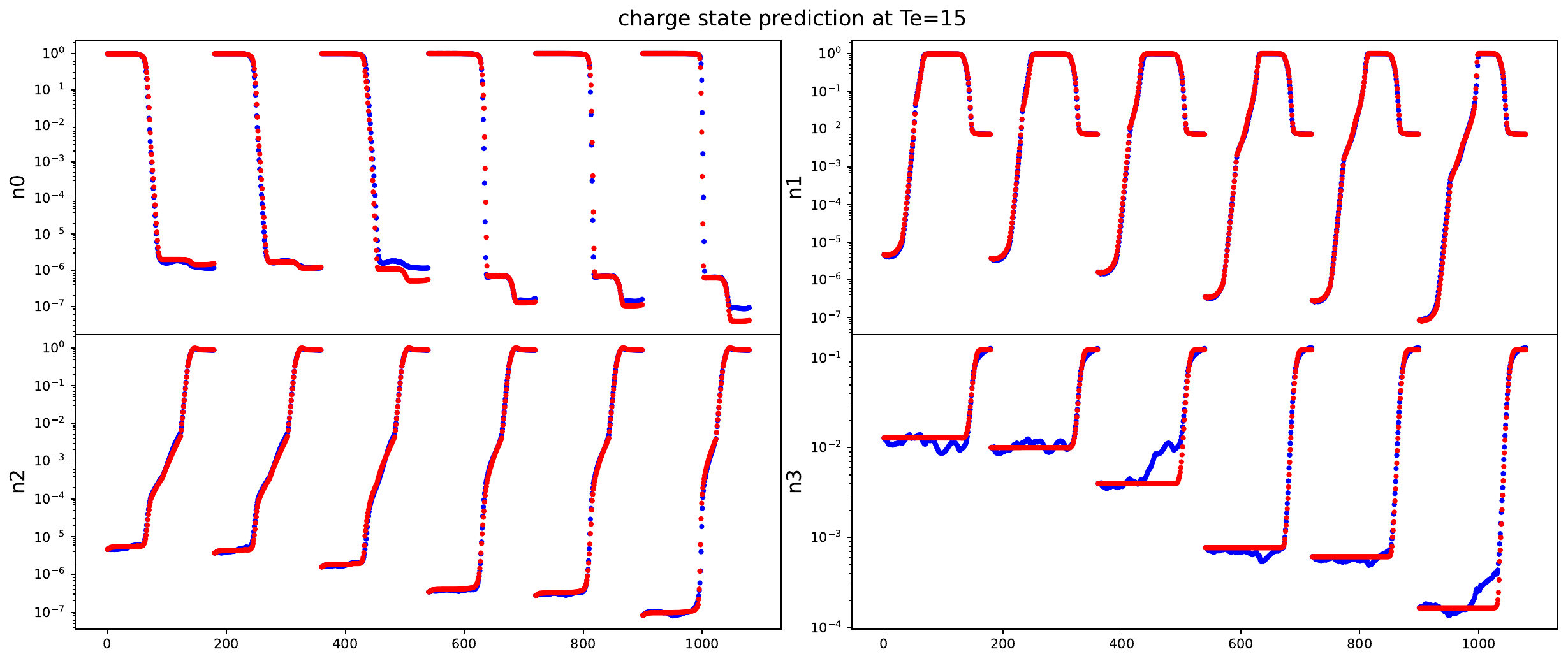}
    \caption{Charge state prediction from the FMNet trained with full $T_e$ values data.}
    \label{fig:pred_4thdata}
\end{figure}

\begin{figure}[!htb]
\centering
        \includegraphics[width=1.0\linewidth]{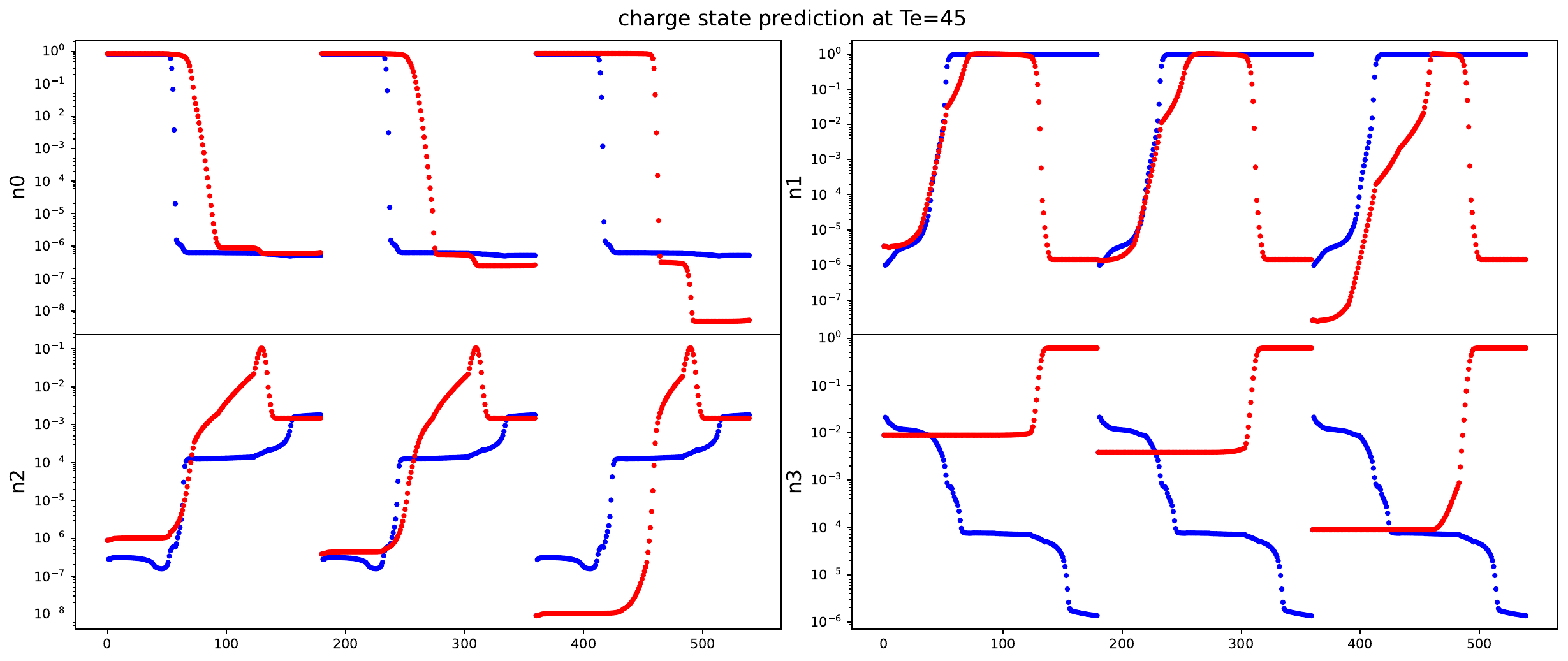}
    \caption{\reva{Charge state prediction from the FMNet trained with first training $T_e$ data (see dark dots in Fig.~\ref{fig:2DTe_values_test}). }}
    \label{fig:pred_1stdatate45}
 \end{figure}

\begin{figure}[!htb]
\centering
        \includegraphics[width=1.0\linewidth]{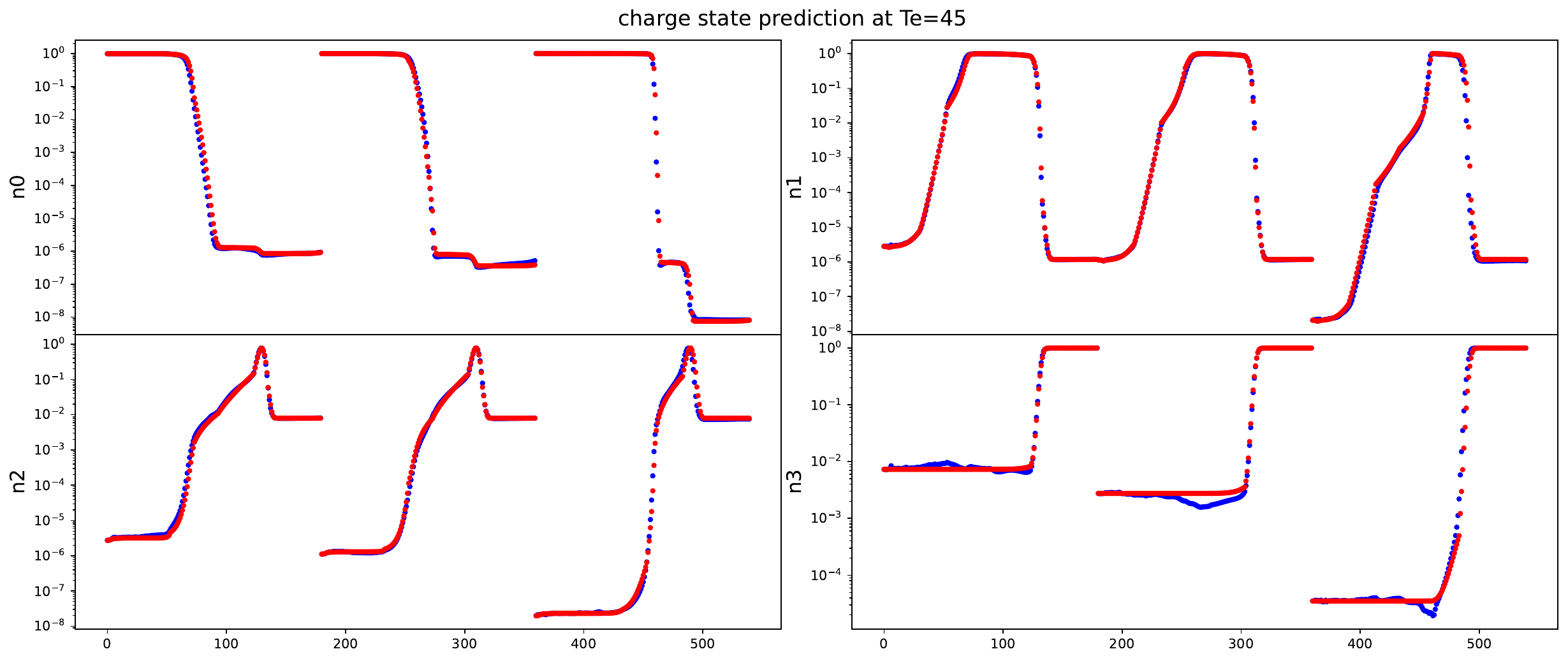}
    \caption{\reva{Charge state prediction from the FMNet trained with full $T_e$ values data.}
    \label{fig:pred_4thdatate45}}
\end{figure}

One of the fundamental principles in ML is that larger datasets tend to produce better models. When training a neural network on a small dataset, the model may not have enough examples to learn robust patterns and relationships in the data. As a result, the model may suffer from overfitting, where it memorizes the training data rather than generalizing well to unseen data. By increasing the size of the training dataset, we provide the model with more diverse examples to learn from, which can help improve its ability to generalize. As the amount of training data increases, the model becomes more exposed to different variations and nuances present in the data, allowing it to learn more robust representations. Consequently, we typically observe better prediction performance as the size of the training dataset grows. Since our dataset is parameterized by the $n_A$ and $T_e$,  we use $T_e$ as the benchmark to test the impact of the datasize. We split the dataset according to the $T_e$ values. For the testing dataset, we use $T_e=$[15, 45, 75, 95]. Our initial training only contains data with $T_e=$[5, 25, 35, 55, 65, 85] (black dots in Fig.~\ref{fig:2DTe_values_test}). The second training we use the full dataset (blue dots in Fig.~\ref{fig:2DTe_values_test}). \reva{Fig.~\ref{fig:pred_1stdata} -- Fig.~\ref{fig:pred_4thdatate45} plot the charge state prediction at $T_e=15$ and $T_e=45$. The results clearly indicate that increasing the training dataset can lead to better prediction performance for the model.}

\section{Conclusion and Future Work}
In this paper, we have introduced a physics-assisted surrogate model
framework tailored for Collisional-radiative (CR) modeling. Our
approach leverages a mixed latent space, comprising a ``white space''
that provides the physical variables for coupling with the plasma
models, and a ``black space'' discovered through an autoencoder to
ensure accuracy.  Subsequently, neural networks are utilized to learn
the dynamics governing this latent space. In the numerical
experiments, by thoroughly evaluating the model's performance under
various initial conditions and parameter settings, we demonstrate its
reliability and effectiveness in predicting the complex dynamics of
the CR model. This comprehensive analysis provides confidence in the
model's applicability to real-world scenarios, where accurate and
efficient predictions are essential for understanding and mitigating
plasma disruptions in fusion reactors. The presented numerical results
substantiate the effectiveness of our approach, demonstrating
promising accuracy in modeling the CR problem.

In future work, we plan to expand our model to include data from
multiple species. Incorporating a broader range of species will
enhance the model's applicability and robustness, allowing for more
comprehensive predictions of radiative loss rates and charge state
dynamics across different plasma conditions. This expansion will
inevitably increase the dataset size and complexity, presenting new
challenges in terms of computational requirements and training costs
which we will use distributed training with multiple
GPUs. Additionally, we aim to integrate NODEs into our
framework. NODEs offer a powerful approach for modeling
continuous-time dynamics, allowing the model to learn the system's
evolution directly from data.


\nocite{*}

\section*{Acknowledgement}
This work was supported by the AI/ML program of the U.S.~Department of Energy (DOE) 
Office of Fusion Energy Science (FES).
QT was also partially supported by the Mathematical Multifaceted Integrated Capability Center (MMICC) and Data-intensive Scientific Machine Learning programs of DOE Advanced Scientific Computing Research (ASCR).

\bibliographystyle{unsrt}
\bibliography{ref}

\end{document}